%% file: main.tex
\documentclass[%
 aip,
 amsmath,amssymb,
preprint,%
 reprint,%
]{revtex4-1}

\usepackage{graphicx}
\usepackage{dcolumn}
\usepackage{bm}

\usepackage[utf8]{inputenc}
\usepackage[T1]{fontenc}
\usepackage{mathptmx}
\usepackage{etoolbox}

\usepackage[english,shorthands=off]{babel}
\usepackage[utf8]{inputenc}
\usepackage{wrapfig}
\usepackage{graphicx}
\usepackage{float}
\usepackage{subfloat}
\usepackage{adjustbox}
\usepackage[caption=false]{subfig}
\usepackage{booktabs} 
\usepackage[table]{xcolor}
\usepackage{rotating}
\usepackage{amsmath}
\usepackage{hyperref}

\usepackage[normalem]{ulem}
\newcommand{\stkout}[1]{\ifmmode\text{\sout{\ensuremath{#1}}}\else\sout{#1}\fi}

\newcommand*{\model}{ANNaMo}

\begin{document}

\preprint{AIP/123-QED}

\title[\model{}: Coarse-grained modelling for folding and assembly of RNA and DNA systems]{\model{}: Coarse-grained modelling for folding and assembly of RNA and DNA systems}
\author{F. Tosti Guerra}
\email[]{francesco.tostiguerra@uniroma1.it}
\affiliation{%
Department of Physics, Sapienza University of Rome
}%
\author{E. Poppleton}
\affiliation{ 
School of Molecular Sciences and Center for Molecular Design and Biomimetics, The Biodesign Institute, Arizona State University
}
\affiliation{Biophysical Engineering Group, Max Planck Institute for Medical Research}
\author{P. Šulc}%
\affiliation{ 
School of Molecular Sciences and Center for Molecular Design and Biomimetics, The Biodesign Institute, Arizona State University
}
\affiliation{Department of Bioscience, School of Natural Sciences, Technical University Munich}
\author{L. Rovigatti}
\email[]{lorenzo.rovigatti@uniroma1.it}
\affiliation{%
Department of Physics, Sapienza University of Rome
}%

\date{\today}

\begin{abstract}
\begin{quotation}
The folding of RNA and DNA strands plays crucial roles in biological systems and bionanotechnology. However, studying these processes with high-resolution numerical models is beyond current computational capabilities due to the timescales and system sizes involved. In this article, we present a new coarse-grained model for investigating the folding dynamics of nucleic acids. Our model represents 3 nucleotides with a patchy particle and is parametrized using well-established nearest-neighbor models. Thanks to the reduction of degrees of freedom and to a bond-swapping mechanism, our model allows for simulations at timescales and length scales that are currently inaccessible to more detailed models.
To validate the performance of our model, we conducted extensive simulations of various systems: We examined the thermodynamics of DNA hairpins, capturing their stability and structural transitions, the folding of an MMTV pseudoknot, a complex RNA structure involved in viral replication, and also explored the folding of an RNA tile containing a k-type pseudoknot. Finally, we evaluated the performance of the new model in reproducing the melting temperatures of oligomers and the dependence on the toehold length of the displacement rate in toehold-mediated displacement processes, a key reaction used in molecular computing.
All in all, the successful reproduction of experimental data and favorable comparisons with existing coarse-grained models validate the effectiveness of the new model.
\end{quotation}

\end{abstract}

\maketitle

\section{Introduction}
\label{introduction}
RNA and DNA molecules play critical roles in biological systems and have become increasingly important in constructing nanoscale architectures \cite{B602886C,dey2021dna}. Successful bionanotechnology designs include DNA origami 
 \cite{rothemund2006folding}, a nanostructure consisting of a long (about 7000 bases) scaffold single-stranded DNA strand, which is compacted into the target shape by designed shorter staple strands that connect different regions on the scaffold. More recently, DNA bricks \cite{ke2012three} consisting just of short strands have been shown to assemble both into 2D and 3D target shapes. Finally, single-stranded DNA and RNA origami and tile structures \cite{qi2018programming,han2017single,geary2021rna} have been designed and shown to be able to also fold into 2D or 3D shapes. These designs consist of a single strand of length ranging from hundreds to thousands of bases, where different regions are complementary to each other, designed to fold into a compact structure consisting of duplex regions and crossovers.  DNA and RNA nanostructures have found a range of applications, ranging from material science to biomedicine \cite{dey2021dna}.
 
 The nucleic acid nanostructure designs are, for the most part, based on DNA and RNA thermodynamics, where the target design maximizes the number of Watson-Crick (and wobble) base pairs present in the system.  However, understanding the kinetics of the folding processes of these designs is essential for unraveling their assembly mechanism and optimizing yields, since kinetic traps can prevent access to the conformations corresponding to global free-energy minima. Access to effective simulation of the folding pathway will facilitate the design of new and more complex nanoscale structures, with applications ranging from nanomanufacturing of plasmonic and photonic devices, molecular robotics, and computing, to the design of more sophisticated tools for biomedical diagnostics and therapeutics.
 
 An additional area of interest is the folding of nucleic acids in biological systems, such as single-stranded genomes of viruses, contranscriptional folding of nascent RNA \cite{watters2016cotranscriptional}, as well as folding of functional RNA molecules, such as designed mRNA for optimized vaccine applications \cite{zhang2023algorithm}. Understanding folding pathways enables the rational design of functional RNA structures with desired properties, such as enhanced catalytic activity or improved binding affinity to specific targets \cite{du2023nucleic,benenson2009rna}. 
 
Computational models have been developed to explore the processes involved in RNA and DNA folding. These models range from atomistic to coarse-grained resolutions, each offering unique advantages and insights into the dynamics of folding. The all-atom model is a commonly employed approach that explicitly represents individual atoms and their interactions. Atomistic force fields such as AMBER \cite{cornell1995second} and CHARMM \cite{brooks1983charmm} have been extensively used to investigate the folding of nucleic acids \cite{sponer2018rna,zgarbova2015refinement}. While these models accurately capture atomistic details and provide valuable insights into the structural and energetic aspects of RNA and DNA, their simulations are computationally demanding and offer limited access to the timescales of interest.

To overcome the limitations of all-atom models, coarse-grained models, which simplify the representation of nucleic acids by grouping multiple atoms into a single particle or bead, have emerged as powerful tools for studying longer lengths and timescales. Several coarse-grained models have been specifically developed for studying DNA and RNA \cite{parisien2008mc,savelyev2009molecular,paliy2010coarse,pasquali2010hire,ouldridge2011structural,10.1063/1.4754132,hinckley2013experimentally,cragnolini2013coarse,xia2013rna,denesyuk2013coarse,10.1063/1.4881424,korolev2014coarse,maciejczyk2014dna,maffeo2014coarse,https://doi.org/10.1002/jcc.23763,machado2015exploring,uusitalo2015martini,dans2016multiscale,ivani2016parmbsc1,chakraborty2018sequence,maffeo2020mrdna,DeLuca2023.06.20.545758}.

Coarse-grained models typically treat the solvent and solvated ions implicitly and represent groups of atoms in the DNA/RNA structure with effective interactions. This simplification enables the study of larger and more extended molecular systems for longer times. Coarse-graining techniques can be broadly categorized as either bottom-up or top-down, each aiming to capture specific aspects of the system. The bottom-up approach formally maps the statistical behavior of a more detailed model into a coarse-grained description, while the top-down approach aims to reproduce as many experimentally relevant properties as possible \cite{sengar2021primer}.

Many of the models mentioned above are parametrized using nearest-neighbor (NN) models. Initially introduced by Poland and Scheraga to investigate duplex denaturation phase transitions \cite{poland1966occurrence}, this approach has been meticulously developed in subsequent years  \cite{doi:10.1146/annurev.biophys.32.110601.141800,Allawi1998,allawi1998thermodynamics,allawi1998nearest2,allawi1997thermodynamics,peyret1999nearest,turner2010nndb,huguet2010single,bae2020high,zuber2022nearest} to describe binding equilibria for oligonucleotides.
Nearest-neighbor models calculate the free energy change ($\Delta G$) associated with forming a duplex by summing the contributions of individual base pair interactions. These interactions are characterized by experimentally determined enthalpy ($\Delta H_{ij}$) and entropy ($\Delta S_{ij}$) values for each possible combination of adjacent base pairs $(i, j)$. Different base pair combinations have different thermodynamic parameters, reflecting variations in hydrogen bonding, stacking interactions, and other factors.

In a NN framework, the binding equilibrium between two isolated strands $A$ and $B$ and their associated fully-bound duplex product $AB$ is fully characterised by the equilibrium constant ($K$), which is defined as the ratio of the concentration of the product $[AB]$ to the concentrations of the individual strands $[A]$ and $[B]$:

\begin{equation}
K = \frac{[AB]}{[A][B]} \propto \exp{(-\beta(\Delta H - T\Delta S))},
\end{equation}
where $\Delta H$ and $\Delta S$ are the total enthalpy and entropy change upon binding, and $\beta=1/k_BT$. By considering nearest-neighbor interactions along the entire sequence length, the model predicts the stability of duplexes and provides insights into their melting temperatures, binding affinities, and overall thermodynamic properties.

In our study, we introduce a new coarse-grained description of nucleic acids where a single patchy particle represents, in principle, $n$ nucleotides, enabling efficient exploration of RNA and DNA folding processes. We parametrize the model using well-established nearest-neighbor models \cite{doi:10.1146/annurev.biophys.32.110601.141800,Allawi1998,allawi1998thermodynamics,allawi1998nearest2,allawi1997thermodynamics,peyret1999nearest,zuber2022nearest} for DNA and RNA thermodynamics to capture essential interactions involved in folding dynamics while maintaining computational efficiency. The new model, named ANNaMo (Another Nucleic-acid Nanotechnology Model), combines the strengths of existing coarse-grained and patchy particle models, providing new possibilities for investigating folding phenomena at previously inaccessible time- and length-scales. This first iteration of the model has been parametrized by fixing $n = 3$, and its thermodynamic performance has been compared with available experimental and numerical data.

\section{Model description}
\label{Model description}
\begin{figure}[!ht]
\begin{tabular}{cc}
    \begin{minipage}{0.27\textwidth} 
    \subfloat[\label{subfig: sketch}]{\includegraphics[width=0.95\textwidth]{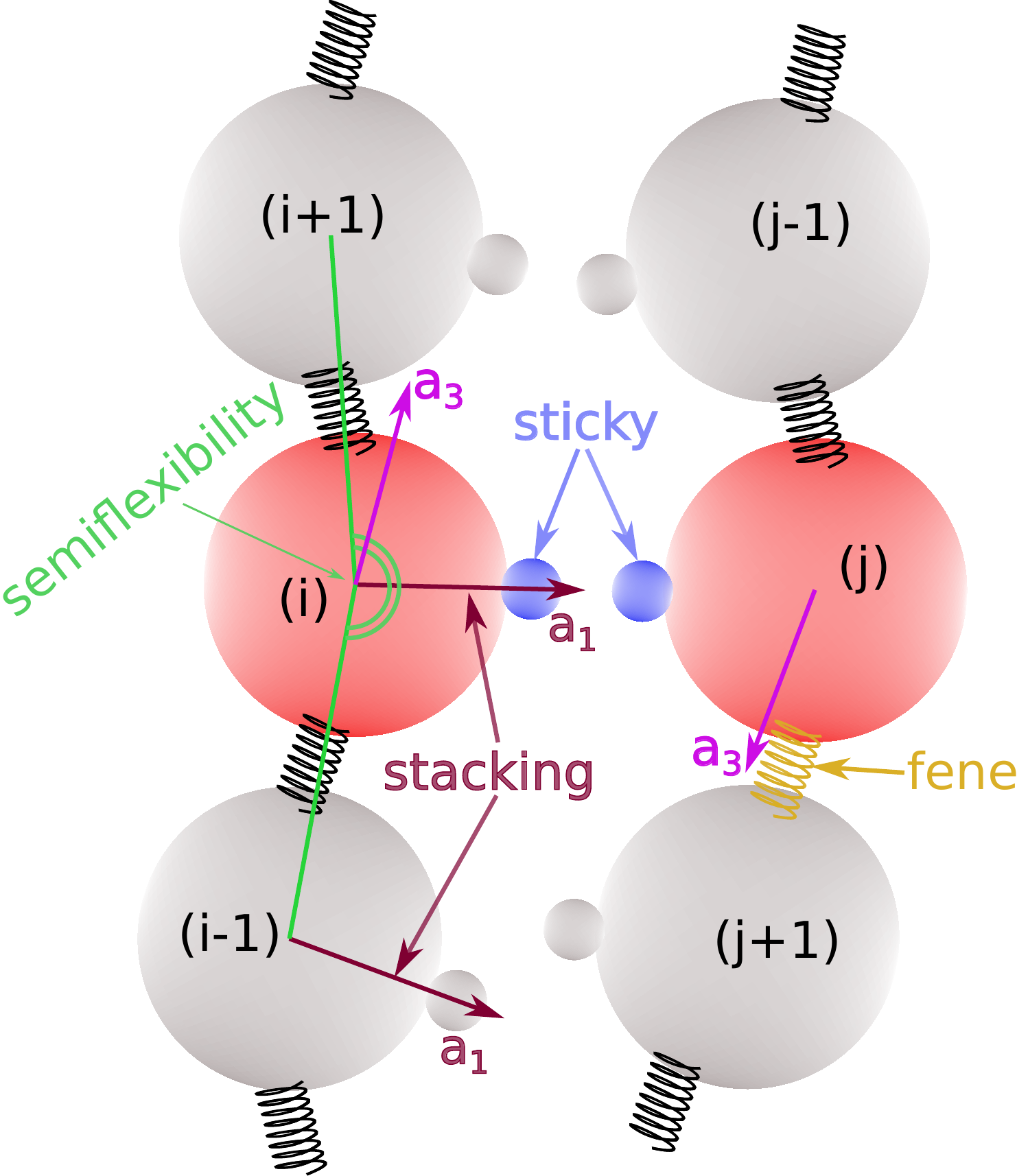} }
    \end{minipage}& 
    \begin{minipage}{0.19\textwidth} 
    \subfloat[\label{subfig: mapping}]
    {\includegraphics[width=0.9\textwidth]{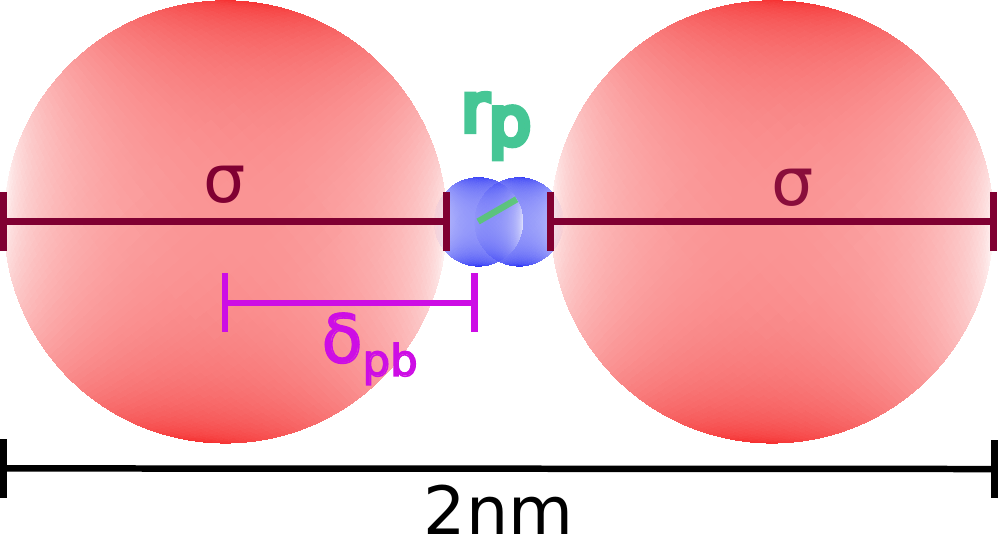}}\end{minipage}
    \end{tabular}
        \caption{\protect\subref{subfig: sketch} schematic representation of the interaction potentials acting between beads in \model{} \protect\subref{subfig: mapping} the diameter of the DNA/RNA helix is approximately 2 nm. Here the bead diameter is $\sigma$, and since we set the distance between the patch and the bead surface to $\delta_{pb} = 0.65\sigma$  and the patch diameter to $r_p = 0.219\sigma$, we can determine that $2\sigma +2\sigma(\delta_{pb} - 0.5)+r_p\sigma=2$ nm. Therefore, we can deduce that $\sigma=2/(1+2\delta_{pb}+r_p)$, or $0.79$ nm.}\label{fig: modelSketch}
    \end{figure}
  
In \model{}, each DNA/RNA strand is modelled as a polymer composed of N beads with diameter $\sigma$. Each bead is decorated with an attractive spherical patch, having a radius of 0.219$\sigma$ and positioned $\delta_{pb}=0.65\sigma$ away from the center of the bead, and represents $n$ consecutive nitrogenous bases. We incorporate several potentials to reproduce the thermodynamic and mechanical characteristics of these biopolymers.

Figure \ref{fig: modelSketch} provides a visual summary of the notation and different interaction terms used in our model, while the detailed functional forms are reported in Appendix~\ref{app:model}. The potentials between topologically bonded beads linked through the backbone include the Kremer-Grest force field \cite{PhysRevA.33.3628}, combining a Weeks-Chandler-Andersen (WCA) potential to model excluded volume, and a finitely extensible nonlinear elastic (FENE) potential to mimic the covalent bonds along the strand. In addition, we introduce a term to add stiffness to the structure: a three-body potential ($V_{\rm semiflex}$) that tends to align three consecutive beads, resulting in a different persistence length between single and double strands (see \ref{subsec:tuning}). To model the stacking of bases in DNA/RNA, we incorporate a term called $V_{\rm stack}$, which promotes the alignment of the directions of consecutive beads that determine the patch positions by using a cosine-angle potential. 

For the non-bonded interactions, we used the WCA potential to account for excluded-volume interactions. Additionally, we employ a patch-patch interaction potential, based on the functional form proposed by Stillinger and Weber \cite{PhysRevB.31.1954}, to model the hybridization of nucleotides. The strength of the attraction between two beads $i$ and $j$, $\epsilon_{ij}$, depends on the types of nucleotides described by each interacting bead and is computed with a nearest-neighbor (NN) model as described below. To ensure that each bead can bind to only one other bead, we implement a repulsive three-body interaction ($V_{\rm 3b}$) that penalizes the formation of triplets of bonded beads \cite{Sciortino2017}. This repulsive potential compensates for the gain associated with the formation of a second bond and can be tuned to favor bond swapping. More, specifically, the parameter $\lambda$ in $V_{\rm 3b}$ allows to interpolate between the limits of swapping ($\lambda = 1$) and non-swapping ($\lambda \gg  1$) bonds. Finally, the $V_{\rm sticky}$ potential is modulated by a term $V_{\rm direct}$ that takes into account the $\Vec{a}_3$ orientations of the beads to ensure that only antiparallel strands can bind to each other.

In the following, we run molecular dynamics simulations at constant temperature~\cite{russo2009reversible} using the \model{} model implemented in the open-source oxDNA simulation engine \cite{poppleton2023oxdna}. The equations of motion are integrated with a velocity Verlet algorithm with a time step $\Delta t=0.002$ (in internal units). A snapshot of the code and examples linked to the systems investigated in this work are also available online~\cite{tosti_guerra_2024_10619450}.

\subsection{Bead design}
\label{Beads designing}
The loss of resolution caused by dividing a strand into beads of $n$ nucleotides leads to a problem where native motifs (\textit{e.g.} hairpins in natural RNAs or crossovers in origami structures) may contain nucleotides from multiple beads or a motif may occur in the middle of a bead. We do not want, however, to make each bead correspond to a unique interaction, thereby losing the ability to simulate unintended folds caused by competing complementary sequences. To balance these demands, our bead design is based on the native contacts observed in folded or designed structures, with the aim of having the majority of the beads representing the same number of nucleotides, $n$. The optimal number of nucleotides per bead is, in general, determined by the desired level of detail and the specific system at hand. However, for this first iteration of the model we optimise the model parameters for $n = 3$, which in the following will be used as the standard size of the beads. However, we allow for a small fraction of the remaining beads to deviate by one nucleotide, either more or fewer, from the standard length, $n$. As a consequence, the beads are composed of varying numbers of nucleotides, requiring the establishment of a rule for calculating interactions. Specifically, we use the strongest $\Delta G$ found by pairing the shorter bead with all possible subsequences of the same length contained within the longer bead.

In the future, we plan to extend the parametrisation to also support different values of $n$, so that it will be possible to simulate DNA and RNA structures at varying levels of detail.

\subsection{Parametrization} 
\label{Parametrization}
The strength of the sticky interaction between any two beads $i,j$ is controlled by a term $\beta \varepsilon_{ij}$ that depends on the nucleotide sequences present in each bead. This term models the pairing of the nucleotides that compose the two beads by taking into account the Gibbs free energy, multiplied by $\beta$, as provided by the nearest-neighbor model. In the following, we will consider systems simulated at a fixed monovalent salt concentration of $0.5$~M, but any other condition can be considered, as long as the NN model's parameters support it.

In order to properly account for temperature variations, we need to separate the contributions from enthalpy and entropy. This can be achieved by rewriting $-\Delta G/k_BT$ as follows:

\begin{align}\label{eq: dG def}
    -\frac{\Delta G}{k_BT} &= -\left( \frac{\Delta H}{k_BT}-\frac{\Delta S}{k_B} \right) \notag\\
    &=-\left( \frac{\Delta H}{k_BT_{\rm ref}} \frac{T_{\rm ref}}{T} -\frac{\Delta S}{k_B} \right)
\end{align}

\noindent
where $T_{\rm ref} / T$ quantifies the temperature difference between the simulation temperature and the temperature $T_{\rm ref}$ used to estimate the values of $\Delta H, \Delta S, \Delta G$ as reported in NN models. Typically, as it is the case here, $T_{\rm ref} = 37$ $^\circ$C.

\begin{figure}
    \includegraphics[width=0.45\textwidth]{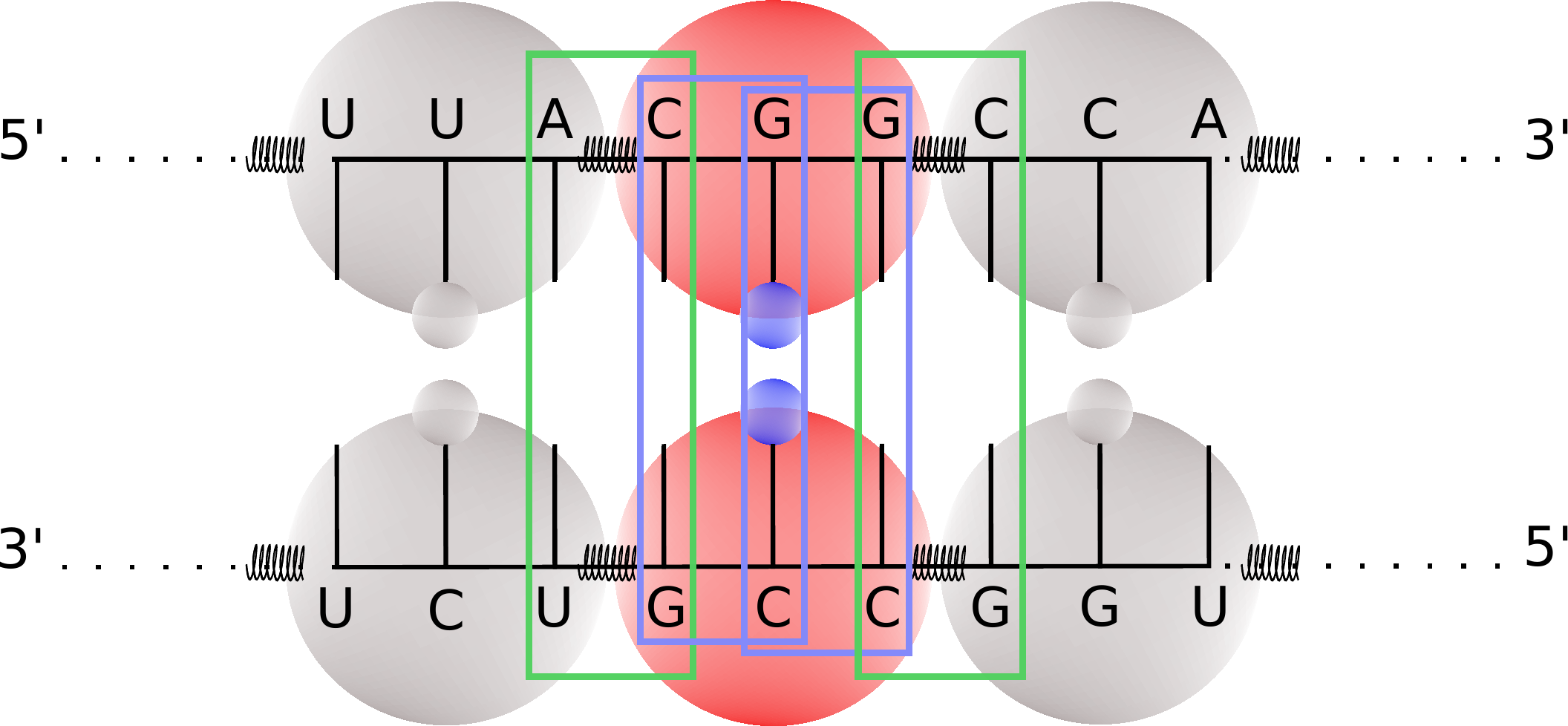}
    \caption{Contributions to the interaction strength between the two colored (red) beads, illustrating the base steps involved. The blue boxes surround base steps made of nucleotides within the beads, while the green boxes surround the two boundary base steps (\textit{i.e.} base steps comprising nucleotides both inside and outside of the beads considered).}
    \label{fig: comb}
\end{figure}

When calculating the $\Delta G$ between two beads, each bead should not be considered in isolation but as part of a larger system, and the coarse-graining procedure should also take into account the nucleotide base steps at the boundaries of the beads. This means including the last nucleotide of the previous bead and the first nucleotide of the following bead for each pair of beads involved in the interaction calculation. However, in order to not overestimate the bead-bead free-energy contribution, the interactions involving nucleotides outside the considered beads are halved. As an example, consider two sequences: $5'-(..A)(CGG)(C..)......((..G)(CCG)(U..)-3'$ (see Figure \ref{fig: comb}). The $\Delta H$ and $\Delta S$ for the interaction between the two middle beads will have the following contributions:
\begin{align}\label{eq: 2beadsInt}
    \begin{pmatrix} 5'-A)(CGG)(C-3' \\[\jot]3'-U)(GCC)(G-5' \end{pmatrix} =\notag\\ \frac{1}{2}\begin{pmatrix} AC \\[\jot]UG \end{pmatrix} + \begin{pmatrix} CG \\[\jot]GC \end{pmatrix} + \begin{pmatrix} GG \\[\jot]CC \end{pmatrix} +\frac{1}{2}\begin{pmatrix} GC \\[\jot]CG \end{pmatrix}.
\end{align}

In order to establish a meaningful correspondence between the potential used in our simulations and the NN parameters, 
the attraction strength between two beads $i$ and $j$, $\epsilon_{ij}$, is computed by equaling the free-energy change due to a bond with that predicted by the NN model, $\Delta H + T \Delta S$~\cite{reinhardt2016dna}, which for the form and parameters of the bead-bead attraction used here gives
\begin{equation}
    \beta \varepsilon_{ij} = -\frac{1}{\alpha}\left[ \left( \frac{\Delta H}{k_BT_{37}} \frac{T_{37}}{T} -\frac{\Delta S}{k_B} \right) +  \ln{(\rho^{\stkout{\circ}}\sigma^3\Tilde{V}_b)} \right],
\end{equation}
where $\rho^{\stkout{\circ}}=[\stkout{\circ}]\mathcal{N}_A=6.022\times 10^{26}\,\textrm{m}^{-3}$ is the standard number density, $\sigma$ is the diameter of the beads and corresponds to 0.79~nm (see Figure \ref{subfig: mapping}), while $\tilde{V}_b = 0.0019 $ and $\alpha= 0.89$ are computed numerically (see Appendix \ref{Mapping} for additional details).

We note that $\epsilon_{ij}$ is the only parameter affected by the type and nature of the nucleotides: in \model{} the geometry and structure of the strands are otherwise independent of the nucleic acid modeled (DNA or RNA) and bead sequence.

We also include the $\Delta H$ and $\Delta S$ associated with the initiation and the terminal penalties estimated by NN models~\cite{doi:10.1146/annurev.biophys.32.110601.141800,zuber2022nearest}. The implementation of the terminal correction is straightforward: given two terminal beads, if the first bead initiates a base step with the last two nucleotides of the second bead, forming a base step tabulated in NN models, we include the associated $\Delta H^{\rm{terminal}}$ and $\Delta S^{\rm{terminal}}$.
By contrast, the initiation term cannot be assigned to two specific beads for each pair of strands, since it is not known beforehand which pair of beads will bond first. As a result, we distribute the initiation contribution evenly among all potential duplex beads: when computing the strength of the interaction between two beads belonging to different strands, we add to it $\Delta H^{\rm{init}}$ and $\Delta S^{\rm{init}}$, as given by NN models, divided by the length (expressed in number of beads) of the shorter strand.

\subsection{Tuning of the parameters}
\label{subsec:tuning}

The free parameters of the model are the strength, denoted as $k_s$, and width parameter, $\xi$, of the $V_{\rm semiflex}$ potential, as well as the stacking strength, $\eta$, of the $V_{\rm stacking}$ potential. We optimize the values of these quantities by comparing them to the melting temperatures of DNA hairpins predicted by the empirically-parameterized oxDNA model and to the persistence lengths of single- and double-stranded DNA, making \model{} a top-down model. For the thermodynamic data, we focus on two sets of hairpins: twelve hairpins with a stem length of six and a loop length of six, and twelve hairpins with a stem length of six and a loop length of nine. The melting temperature is defined as the temperature at which the yield of the closed hairpin is $0.5$, and the hairpin is considered closed when at least 3 nucleotides are bonded for oxDNA simulations, or one bead-bead bond is present for the \model{} model. The estimation of the persistence lengths is carried out by simulating 200-bead hairpins with stems of 98 beads, and 100-bead single strands. In both cases, we define the unit vector $\hat{v}_i$ connecting each pair $i$ and $i + 1$, and define $\cos(\theta_{ij}) = \hat{v}_i \cdot \hat{v}_j$, where $i$ and $j$ are separated by $m$ beads. Averaging over all pairs (excluding the terminal 10 beads), we obtain an angular correlation, $\langle \cos(\theta_{ij}) \rangle$, which is a function of $m$ only. For worm-like chains, such a correlation decays as

\begin{equation}
\langle \cos(\theta_{ij}) \rangle = e^{-m/l_p}.
\label{eq:correlation}
\end{equation}

We apply Eq.~\eqref{eq:correlation} to extract the persistence length from our simulations. The potential we use tends to favor the presence of kinks in double strands, which decreases their persistence length. Therefore, we perform the analysis described above with and without kinks, which are defined as $i,j$ neighboring pairs for which $\cos(\theta_{ij}) < 0.7$. While the proper persistence length requires that all angles are taken into account, excluding kinks provides an estimate of the rigidity of the double-stranded parts of the structures that do not bend much, which is the most common state of double strands in complex structures (\textit{i.e.} origami).

Through extensive analysis, we found that the closest results to the oxDNA predictions and the estimated persistence length for single- and double-stranded DNA are obtained when using $\beta k_s=4$, $\xi=0.07$ and $\beta \eta=6$. The resulting melting temperatures are shown in Table \ref{tab:hair}, comparing very well to those extracted from oxDNA simulations. Indeed, the two data sets differ by an average of just about two kelvins.

\begin{figure}
    \centering
    \includegraphics[width=0.45\textwidth]{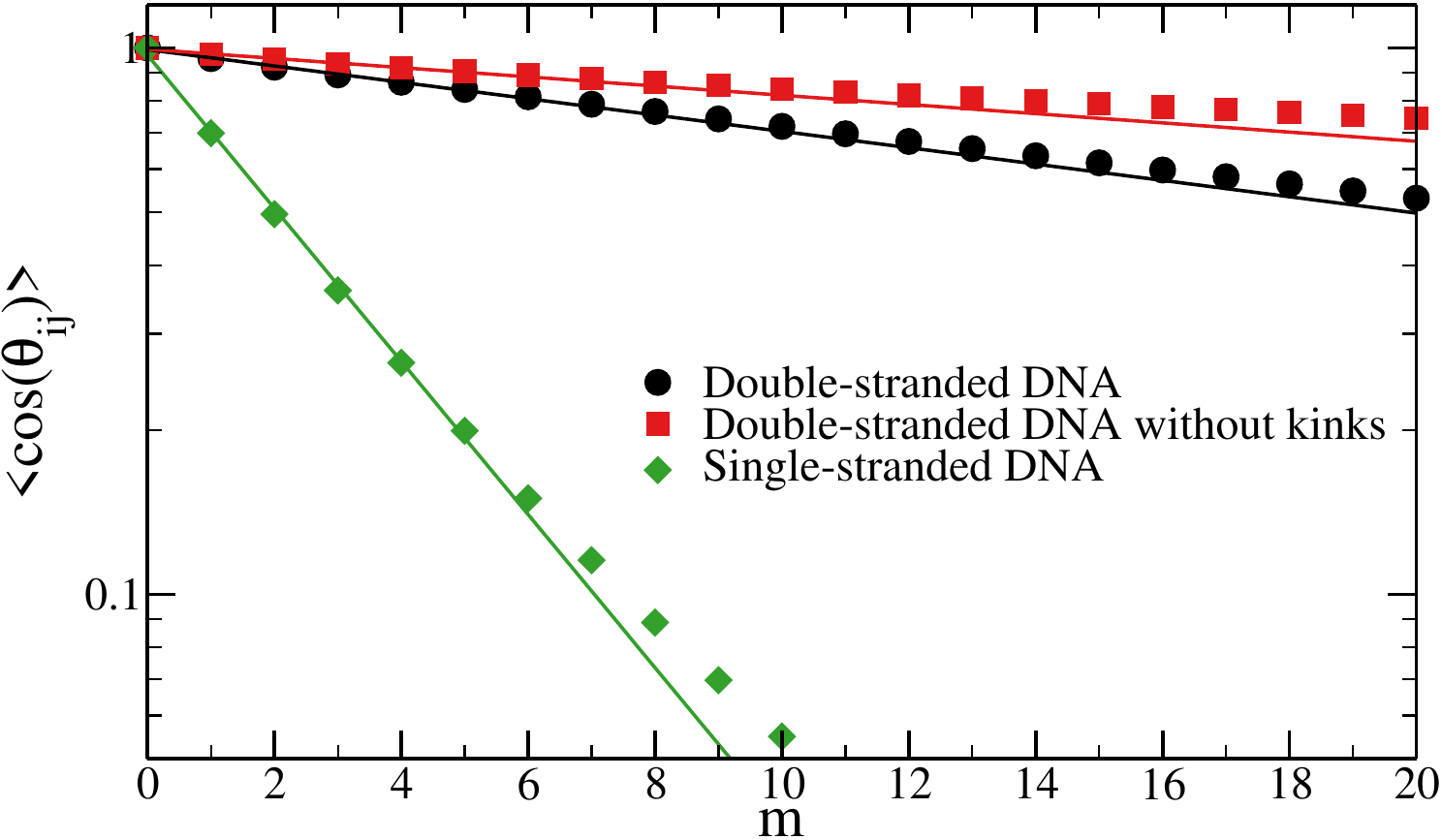}
    \caption{Angular correlation as a function of the chemical distance $m$ for double-stranded and single-stranded DNA molecules. Points are simulation data, lines are exponential fits performed in the $m \in [0, 5]$ interval. Note that, as for real DNA and RNA, single strands do not behave as worm-like chains and therefore the angular correlation decays only approximately as an exponential.}
    \label{fig: lp_ss_ds}
\end{figure}

Figure~\ref{fig: lp_ss_ds} illustrates that the persistence length of the double strand is approximately 28 beads with kinks and 49 without kinks, equivalent to around 84 and 147 base pairs (bp). In contrast, the derived persistence length of single strands is approximately 3 beads, corresponding to 9 bp. These values align well with estimates for real DNA, where the persistence length of double-stranded DNA is approximately 50~nm\cite{wang1997stretching} (equivalent to around 147~bp), and that of single-stranded DNA is approximately $(1.98\pm0.72)$~nm~\cite{roth2018measuring} (equivalent to approximately $(5.8\pm2.1)$~bp), considering a base pair spacing of 0.34~nm.

\begin{table}[h!]
    \input{tables/6stem6loop}

    \bigskip
    \bigskip
    \input{tables/6stem9loop}

    \caption{Melting temperatures of selected hairpins (in °C), as obtained with oxDNA and \model{}. The corresponding sequences are listed in Appendix~\ref{sequences}. $|\Delta|$ is the absolute value of the difference between the oxDNA and \model{} predictions, and $\langle|\Delta|\rangle$ is its average.}\label{tab:hair}
\end{table}

\section{Results}

\subsection{DNA hairpins} \label{sub: DNA hairpins}

\begin{figure}[!ht]
\begin{tabular}{c}
    \begin{minipage}{0.45\textwidth} 
    \subfloat[\label{subfig: loop6}]
    {\includegraphics[width=0.9\textwidth]{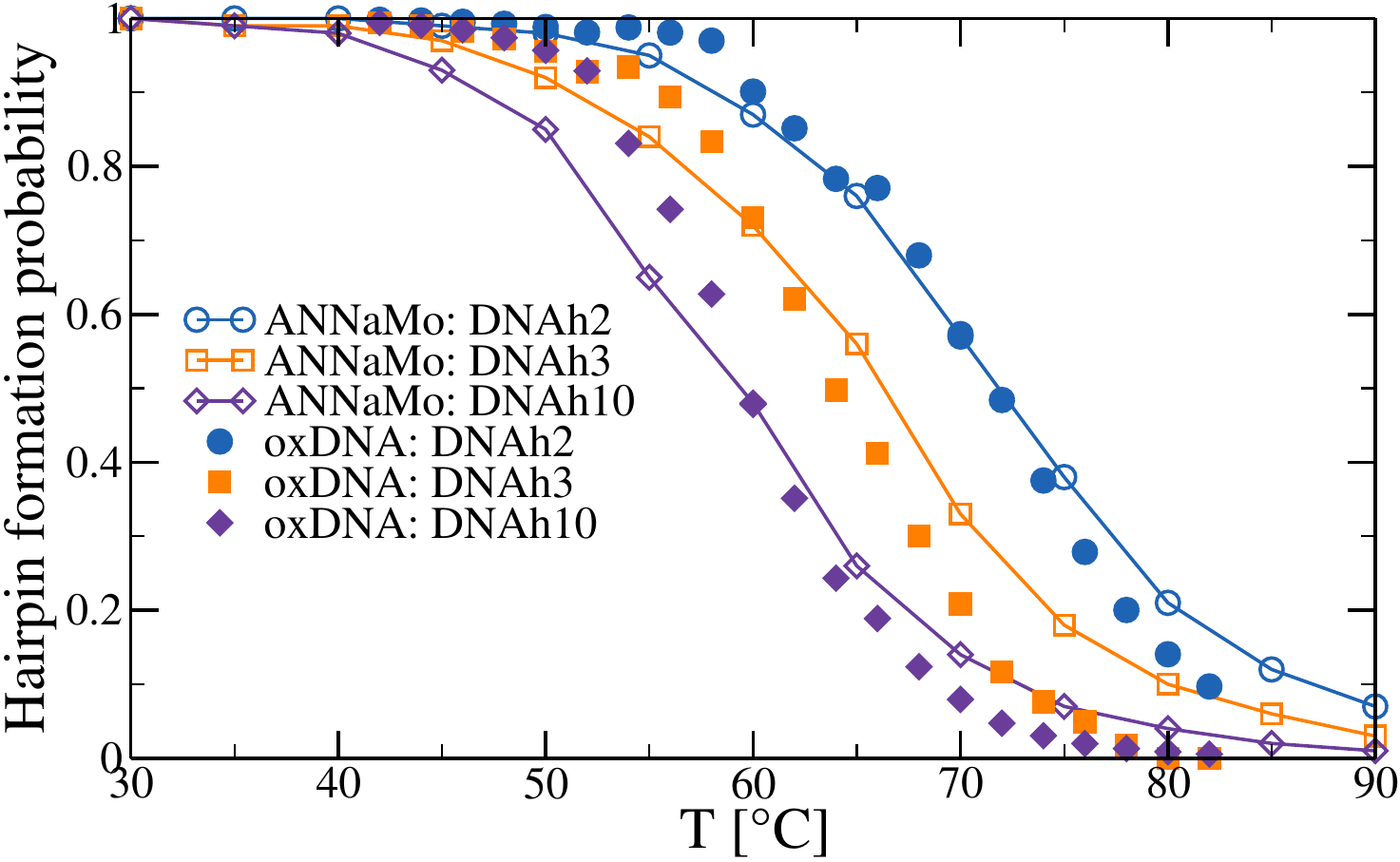} }
    \end{minipage} \\
    \begin{minipage}{0.45\textwidth} 
    \subfloat[\label{subfig: loop9}]
    {\includegraphics[width=0.9\textwidth]{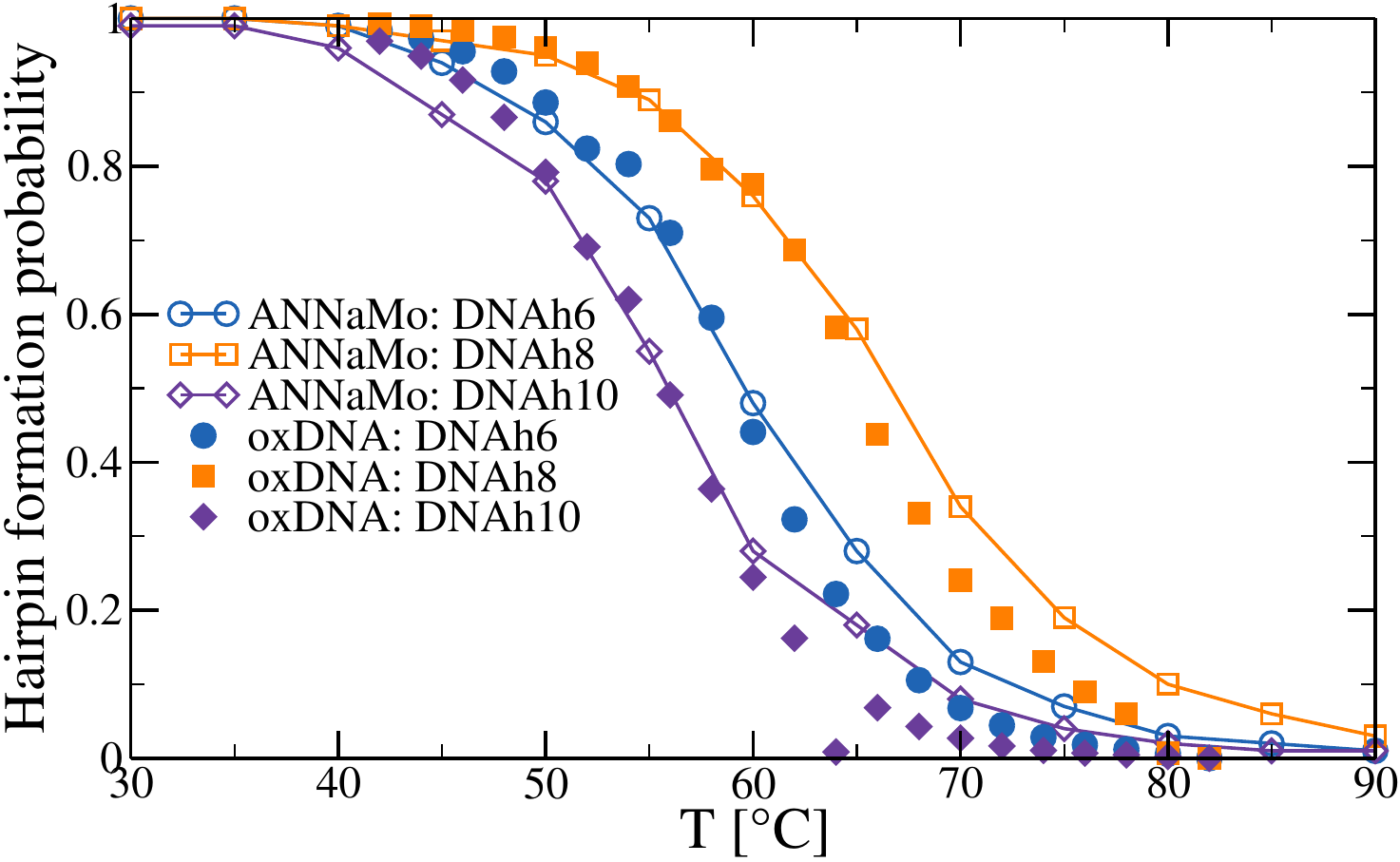} }
    \end{minipage}
    \end{tabular}
        \caption{Comparison of the melting curves predicted by oxDNA (points) and \model{} (lines) for selected DNA hairpins with a stem length of six and a loop length of six \protect\subref{subfig: loop6}, and selected DNA hairpins with a stem length of six and a loop length of nine \protect\subref{subfig: loop9}. The resulting melting temperatures are listed in Table \ref{tab:hair}.}
    \label{fig: hairpinStudy}
    \end{figure}

We start by comparing the melting curves of DNA hairpins simulated with oxDNA and \model{}. A representative set of these results, taken from the dataset used to tune the model parameters (see Table~\ref{tab:hair}), is shown in Figure~\ref{fig: hairpinStudy}. For all the hairpins considered, the agreement is almost quantitative for a large temperature range, with oxDNA hairpins exhibiting a somewhat narrower transition, \textit{i.e.} slightly steeper melting curves. At high temperatures the \model{} seems to fall off more slowly compared to oxDNA: we ascribe this behavior to the fact that the bonding volume of bead-bead interactions is rather large, so that there is always a sizeable chance that two beads are considered to be bonded for purely geometrical reasons even for temperatures at which the attraction strength is small compared to the thermal energy.

We use hairpin systems to also estimate the performance difference between oxDNA and \model{}. Specifically, we run single-core molecular dynamics simulations of hairpins at the melting temperature and count the number of transitions between the two states (open and closed). Dividing the resulting number by the simulation wall time yields a factor of $\approx 100$, which we use to estimate the sampling speed-up of the new model compared to oxDNA. While the precise figure should (and will) depend on the system at hand, it is reasonable to expect the speed-up to be of the same order of magnitude, independently on the specific conditions.

\subsection{RNA pseudoknot}
A pseudoknot is a folding pattern that involves the formation of additional base pairs between distant regions of the RNA sequence, resulting in a knot-like structure. Pseudoknots are known to have crucial functions in various biological processes such as regulation of gene expression and ribosome function \cite{peselis2014structure}, and viral replication\cite{neupane2021structural}. They have also been proposed to have a notable influence on RNA folding pathways \cite{10.1093/bioinformatics/btv572}.
The thermodynamics of pseudoknot base-pairing remains poorly characterized due to their enormous diversity and complexity. Furthermore, the fact that they are composed of non-nested base pairs means that they cannot be computed using the efficient dynamic-programming approaches which dominate in the RNA secondary structure prediction field\cite{eddy2004rna}. Therefore, exact and heuristic coarse-graining methods have mainly focused on pseudoknot-free secondary structures.
\begin{figure}[!ht]
\begin{tabular}{cc}
    \begin{minipage}{0.1\textwidth} 
    \subfloat[\label{subfig: smartdiv}]{\includegraphics[width=0.95\textwidth]{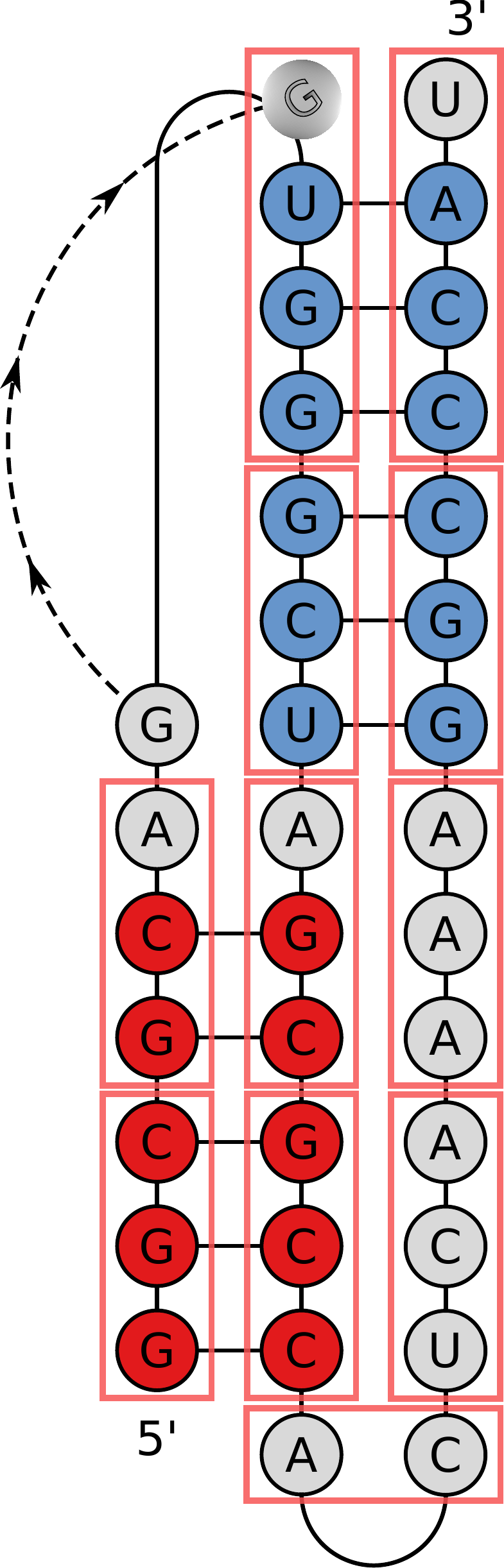} }
    \end{minipage}& 
    \begin{minipage}{0.36\textwidth} 
    \subfloat[\label{subfig: nNxBpse}]
    {\includegraphics[width=1\textwidth]{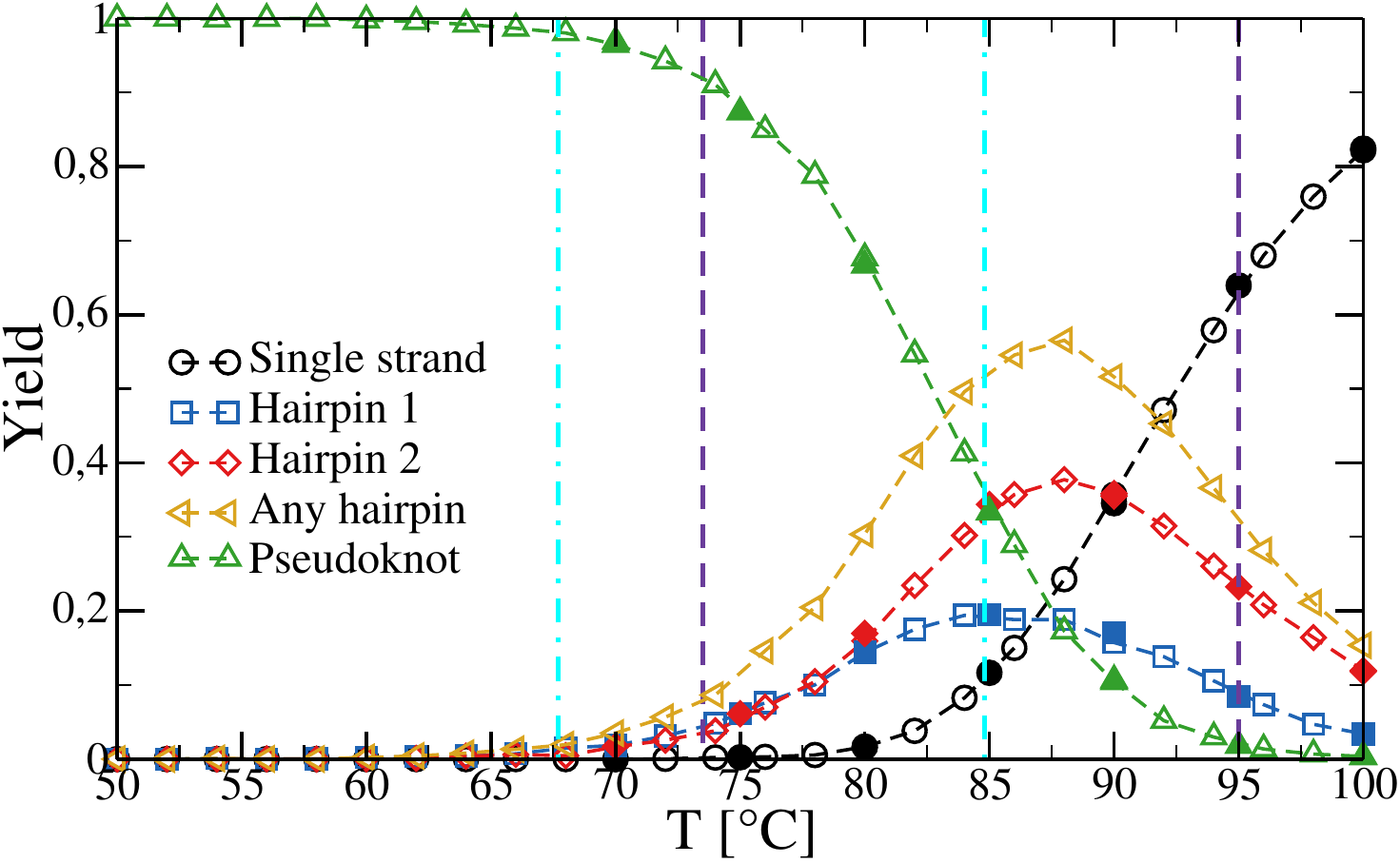}}\end{minipage}
    \end{tabular}
        \caption{\protect\subref{subfig: smartdiv} A schematic representation of the secondary structure of the MMTV pseudoknot, showing the bead division used to simulate the strand with \model{}. Nucleotides inside the same box belong to the same bead, while nucleotides outside the box belong to the bead that follows.  
 \protect\subref{subfig: nNxBpse} \model{} equilibrium yields for the MMTV pseudoknot. Filled symbols represent the results of simulations without swapping, while empty symbols represent those with swapping. The transition temperatures determined experimentally~\cite{theimer2000contribution} and numerically, with oxRNA~\cite{10.1063/1.4881424}, are indicated by violet dashed lines and cyan dash-dotted lines, respectively.}\label{fig: pseudo}
    \end{figure}

We test our model by exploring the melting curve of the well-known MMTV pseudoknot \cite{theimer2000contribution}. We divide the sequence in beads as depicted in Figure \ref{subfig: smartdiv} and we compare our predictions with experimental calorimetry measurements~\cite{theimer2000contribution} and oxRNA simulations~\cite{10.1063/1.4881424}, in Figure~\ref{subfig: nNxBpse}. The yields of the two pseudoknot-precursor motifs (hairpin 1 and 2, highlighted in blue and red in Figure~\ref{subfig: smartdiv}) display the same qualitative trends observed in oxRNA (reported in Ref.~\cite{10.1063/1.4881424}), although the peaks for hairpin 1 and hairpin 2 are higher and lower in \model{} compared to oxRNA, respectively. Summing up the yields of the two intermediate hairpins we obtain a curve (gold points in the figure) whose intersection with the single strand and pseudoknot data identifies the two transition temperatures, 83~$^\circ$ C and 92~$^\circ$C. Compared to the experimental temperatures (74~$^\circ$C and 95~$^\circ$C), these values are at least as good as the ones obtained with oxRNA simulations (68~$^\circ$C and 85~$^\circ$C) which, given the lower resolution and much higher computational efficiency of the \model{} compare to the latter, is an excellent result. 

We have also run simulations without the swap mechanism, which when enabled serves the purpose of speeding up equilibration and sampling, by setting $\lambda = 10$. As shown in Figure~\ref{subfig: nNxBpse}, the results obtained with and without the swap (empty and full symbols, respectively) overlap perfectly, demonstrating that the thermodynamics is not affected by the value of $\lambda$.

\subsection{RNA tile}

Drawing inspiration from the idea of molecular tiles, which are individual units that self-assemble to create intricate patterns or structures\cite{winfree1998design}, we simulated the assembly of a single-stranded RNA tile containing a k-type pseudoknot\cite{poppleton2020design}. 

Unfortunately, no comparison with oxRNA is possible, as observing the folding of such a structure, composed of 132 nucleotides, is currently out of reach from the computational point of view. As before, we split the sequence into beads of average size 3, but optimize the division by hand so that nucleotides that are supposed to be paired in the native structure belong to the same beads.

  \begin{figure}[!ht]
    \centering
    \begin{tabular}{c}     
    \adjustbox{}{\begin{tabular}{@{}c@{}}
    \subfloat[\label{subfig: Snap}]{%
          \includegraphics[width=.25\textwidth]{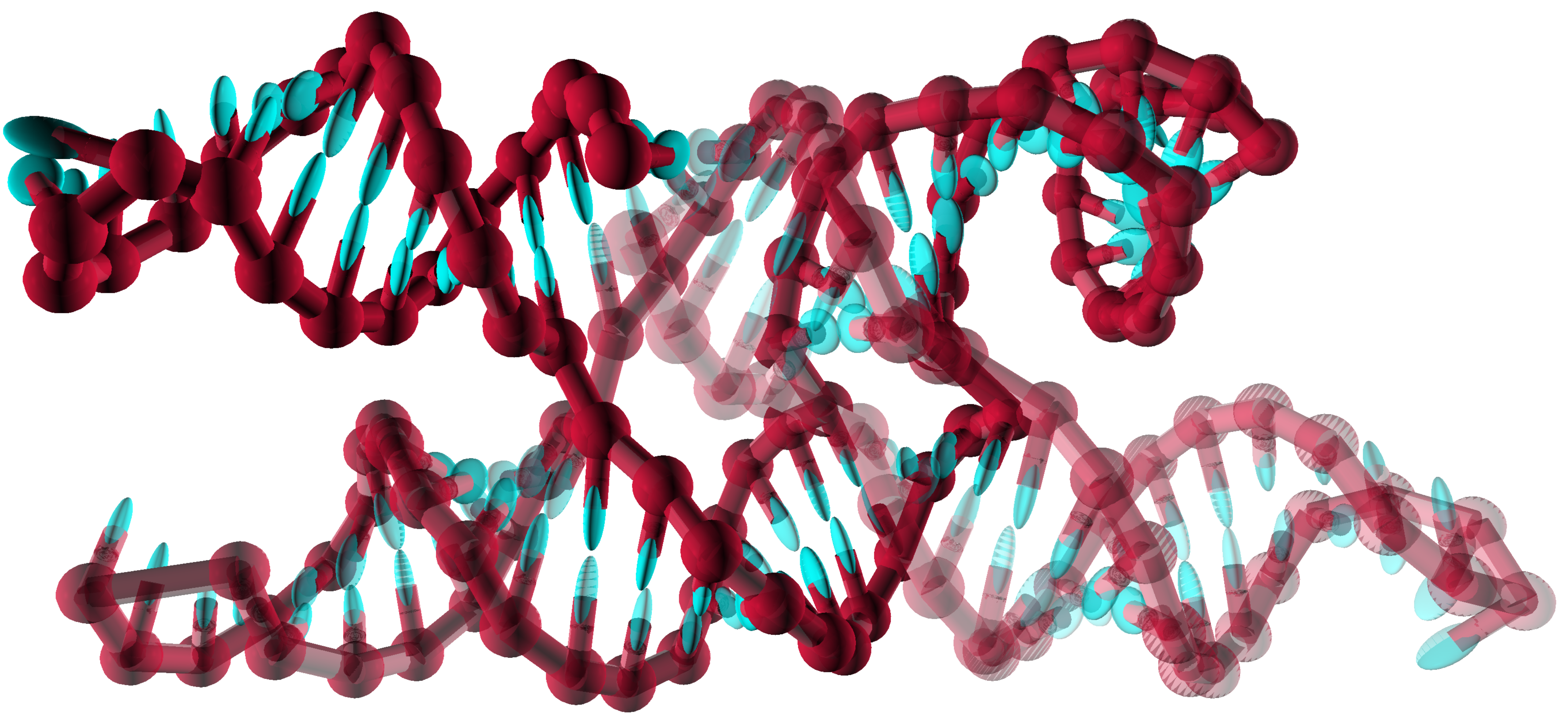}} 
    \subfloat[\label{subfig: LastConf}]{%
          \includegraphics[width=.2\textwidth]{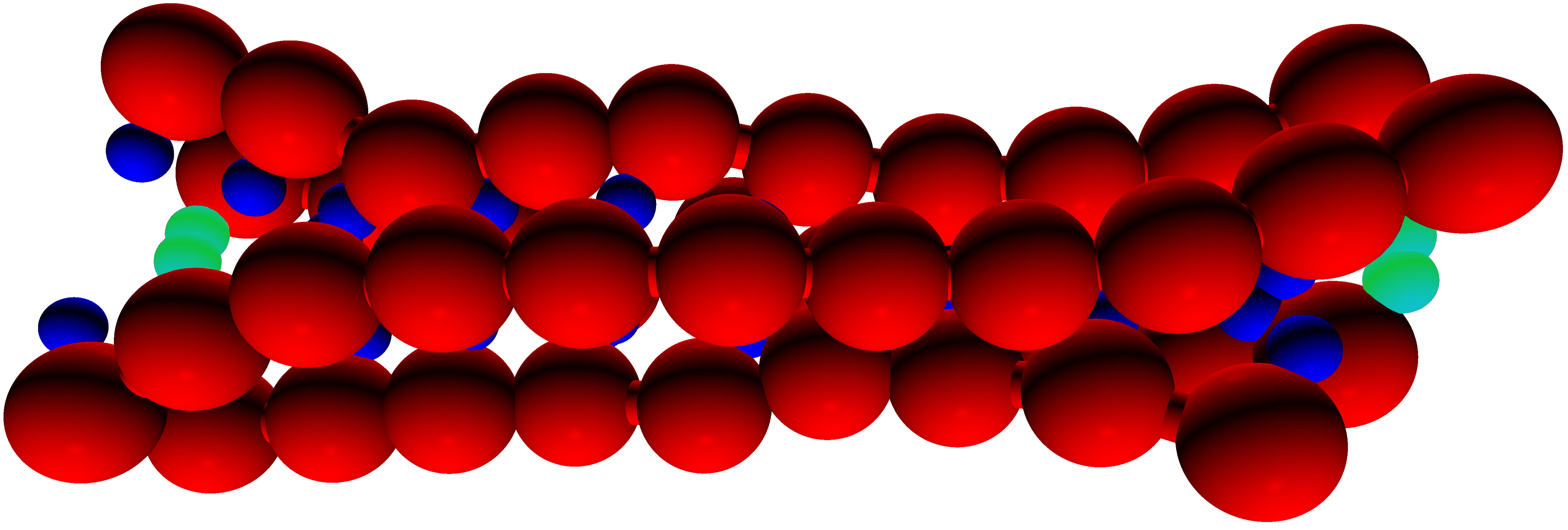}}
    \end{tabular}}
    \\
    \adjustbox{}{\subfloat[\label{subfig: tileMelt}]{\includegraphics[width=.45\textwidth]{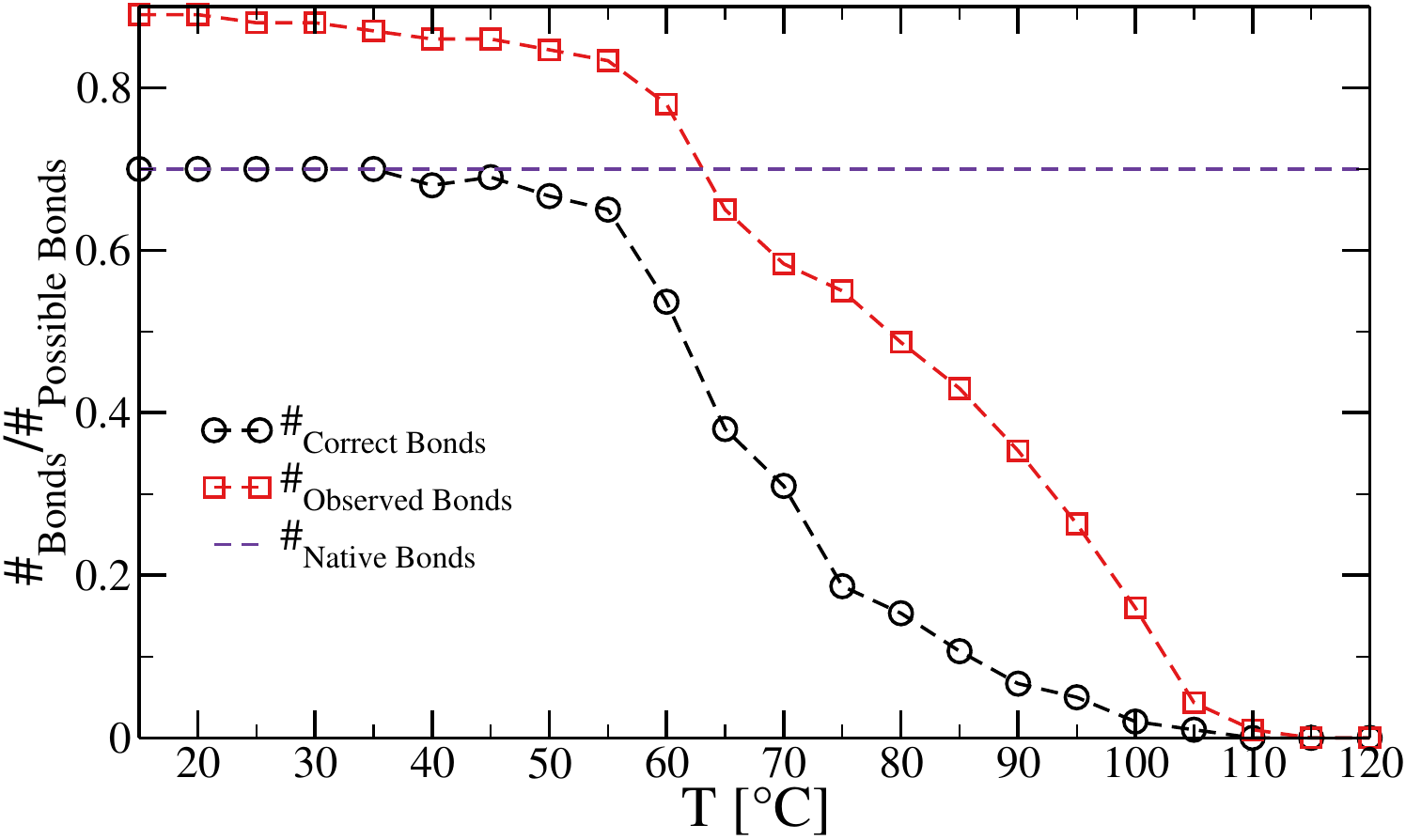}}}
    \end{tabular}
    \caption{\protect\subref{subfig: Snap} The RNA tile composed of 132 nucleotides \cite{poppleton2020design} we simulate, as represented by oxRNA. \protect\subref{subfig: LastConf} A \model{} configuration simulated at $28^\circ$C. The highlighted (green) patches form kissing loops.\protect\subref{subfig: tileMelt} Melting curve of the RNA tile simulated with the \model{} model.}\label{fig: tile}
  \end{figure}

We simulate the tile for different temperatures, evaluating the fraction of bonds between any two beads, and between beads that should be bonded in the native structure. The results are shown in Figure~\ref{fig: tile} and should be compared with the dashed horizontal line that signals the fraction of bonds that are bonded in the native structure ($\approx 0.7$). The \model{} model predicts a melting temperature (\textit{i.e.} a temperature at which half of the native contacts are formed) around 65$^\circ$, and that below $\approx 50^\circ$ nearly all bonds are formed.

\begin{figure}
    \centering
    \includegraphics[width=0.45\textwidth]{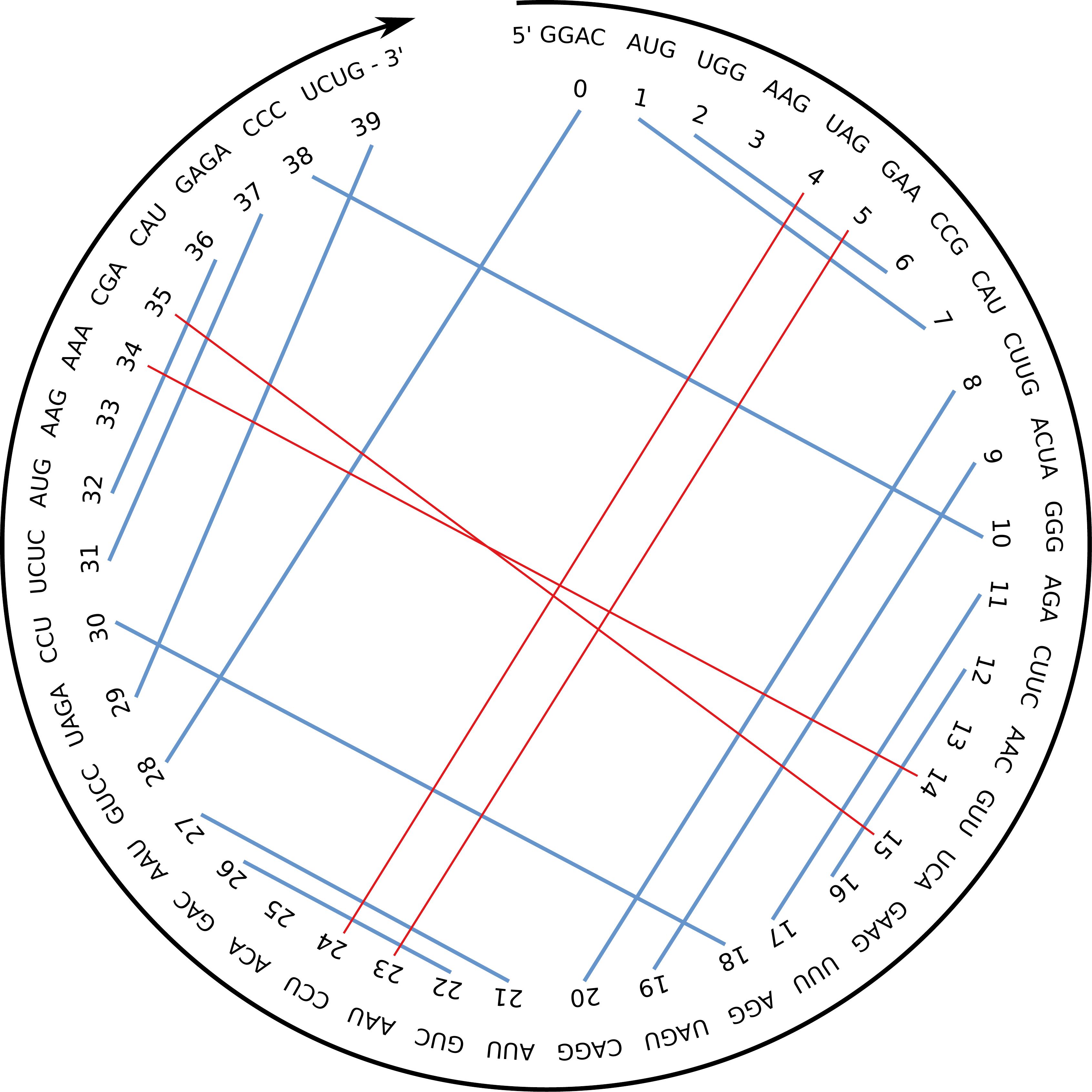}
    \caption{Circular representation of the RNA tile structure and of its sequence, alongside the adopted division into beads (numbers) and the bonds that we observe at low temperature (lines). Blue lines indicate native bonds, while red lines indicate the most probable bonds that connect beads that are not fully complementary but are nevertheless observed in simulations.}
    \label{fig: tileBonds}
\end{figure}

At low temperatures, on top of the native contacts, which are always formed, we observe additional bonds that link beads that are not fully compatible with each other. Figure~\ref{fig: tileBonds} shows the tile sequence, the splitting into beads and the native contacts (blue lines), as well as the most probable misbonds observed (red lines). It is clear that most misbonds happen between beads that are part of hairpin loops, thereby forming so-called kissing loops (highlighted in figure~\ref{subfig: LastConf}). Kissing loops complexes are known to play an important role in RNA-RNA interactions, both in the biological and nanotechnology contexts~\cite{paillart1996loop,bindewald2008rnajunction,liu2020branched}.
Although it is hard to estimate the stability of these motifs within the tile with oxRNA or similar coarse-grained models, the lack of flexibility of the tile arms should disfavour these particular kissing loops. Therefore, it is possible that the \model{} model overestimates their stability. 

\subsection{DNA double strands}
Without further parameter optimizations, we compared the melting temperatures of DNA oligomers with that predicted by the SantaLucia's NN model. We simulated duplexes of varying lengths, ranging from 2 to 8 beads made of 3 nucleotides each, equivalent to 6 to 24 base pairs. As we simulate small systems where a single assembly can form, in order to mitigate finite-size effects resulting from the suppression of concentration fluctuations, we employed the approach introduced in Ref.~\cite{ouldridge2010extracting}. 
\begin{figure}
    \centering
    \includegraphics[width=.45\textwidth]{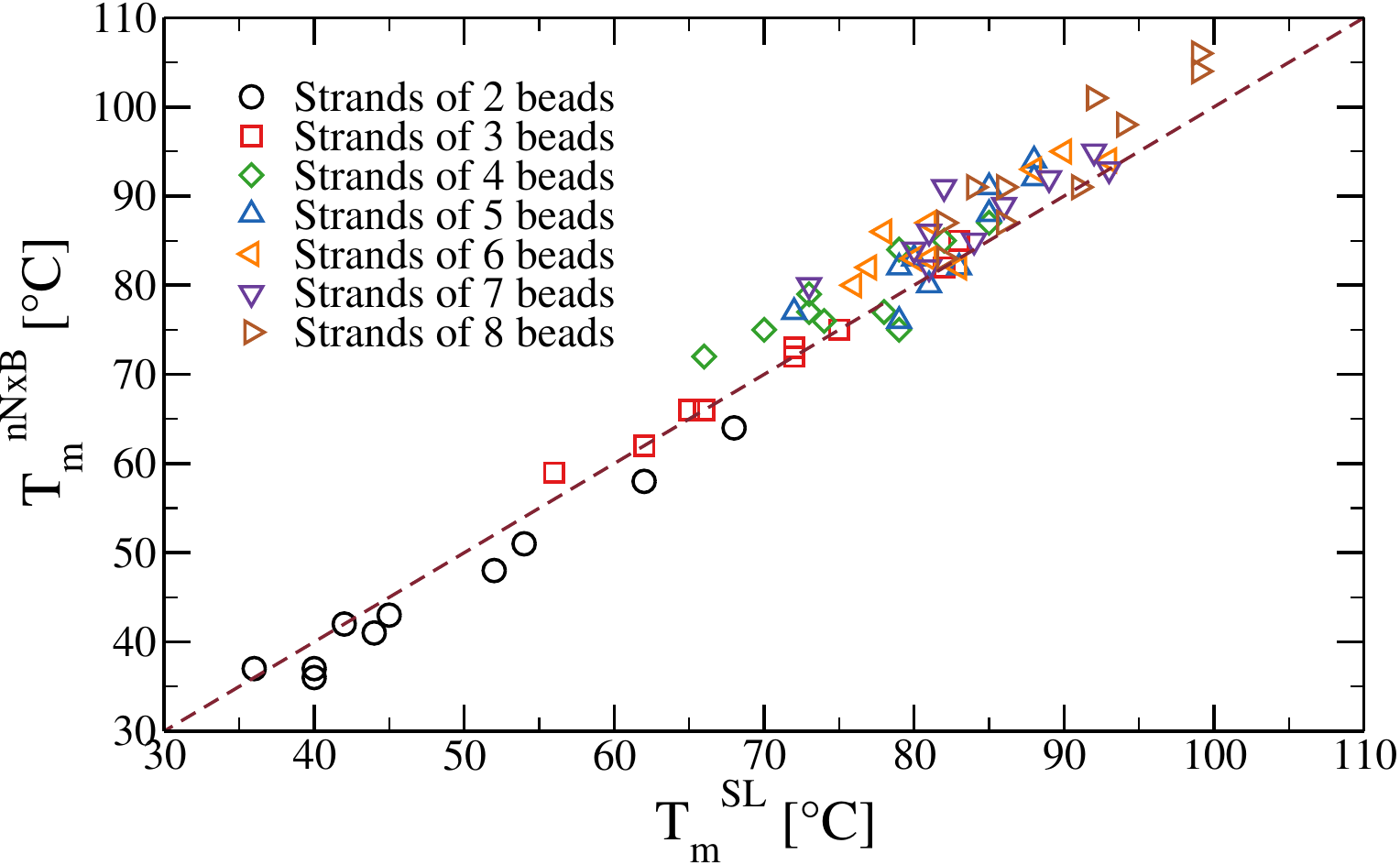}
    \caption{Melting temperatures ($T_m$) predicted by \model{} as a function of the corresponding predictions of the SantaLucia model for 70 duplexes of 6 to 24 bp in length. The average absolute deviation is $\approx 3\,^\circ$C.}
    \label{fig:res_ms}
\end{figure}

The results, reported in Figure~\ref{fig:res_ms}, show that the \model{} melting temperatures are always rather close to those predicted by NN models, with an average absolute deviation of $\approx 3\,^\circ$C, comparable to that observed for the hairpins investigated earlier. The model tends to underestimate melting temperatures of short strands, and overestimate those of longer strands, with oligomers of length 3 and 4 beads (9-12 bp), which is a range of common domain sizes in DNA nanotechnology, performing best.

\subsection{Toehold-mediated strand displacement}

Toehold-mediated strand displacement (TMSD) processes are a key mechanism used in molecular computing \cite{simmel2019principles}. It consists of an invader strand that displaces an incumbent strand that was previously bound to a substrate strand. The substrate can be longer than the incumbent strand, resulting in a single-stranded region (toehold) to which the invading strand can bind. The length of the toehold can be used to tune the kinetics of DNA- and RNA-nanotechnology systems~\cite{D0NR07865D}. Especially in larger reaction cascades consisting of many strand species, using the toehold length to fine-tune the kinetics of the individual strand-displacement reactions can be required to achieve good performance of the molecular circuit~\cite{WANG2023107906}. The biophysics of the process has been investigated in detail with both experiments and coarse-grained simulations~\cite{zhang2009control,srinivas2013biophysics,Walbrun2024.01.16.575816}. However, most off-lattice 3D coarse-grained models are too detailed to directly probe TMSD events in an unbiased fashion, and rare-event techniques, such as umbrella sampling or forward-flux sampling (FFS)~\cite{allen2009forward,10.1063/1.5127780}, have to be deployed in order to obtain reliable estimates of thermodynamic and kinetic quantities.

Here we use the three-strand system studied in Refs.~\cite{zhang2009control,srinivas2013biophysics}: a 20-nucleotide-long incumbent strand is complementary (and bound) to a substrate that has an additional toehold of variable length (ranging from 0 to 15 nucleotides); the third strand is perfectly complementary to the substrate. If the toehold is present, then the stable thermodynamic state is the one where the invader is bound to the substrate after having displaced the incumbent. If no toehold is present, the incumbent-substrate and invader-substrate states have the same free energy. The main kinetic quantity of interest is the displacement rate, which can be estimated through experiments, as well as through numerical simulations of coarse-grained models~\cite{zhang2009control,srinivas2013biophysics,Walbrun2024.01.16.575816}. 

Here we use \model{} to evaluate the displacement rate as a function of toehold length (up to $5$ beads, corresponding to $15$ nucleotides) with and without the bond-swapping mechanism. As done in Ref.~\cite{srinivas2013biophysics} for oxDNA, we use an interaction matrix where the only non-zero entries are those relative to the native contacts, and we do not take into account the time spent in three-stranded complexes to evaluate rates in order to make it possible to compare results with experiments. For the swapping case and for the non-swapping cases with non-zero toehold length, the dynamics of \model{} is fast enough that brute-force calculations are possible. In the other cases we resort to performing FFS calculations.

\begin{figure}[!ht]
\centering
      \includegraphics[width=.45\textwidth]{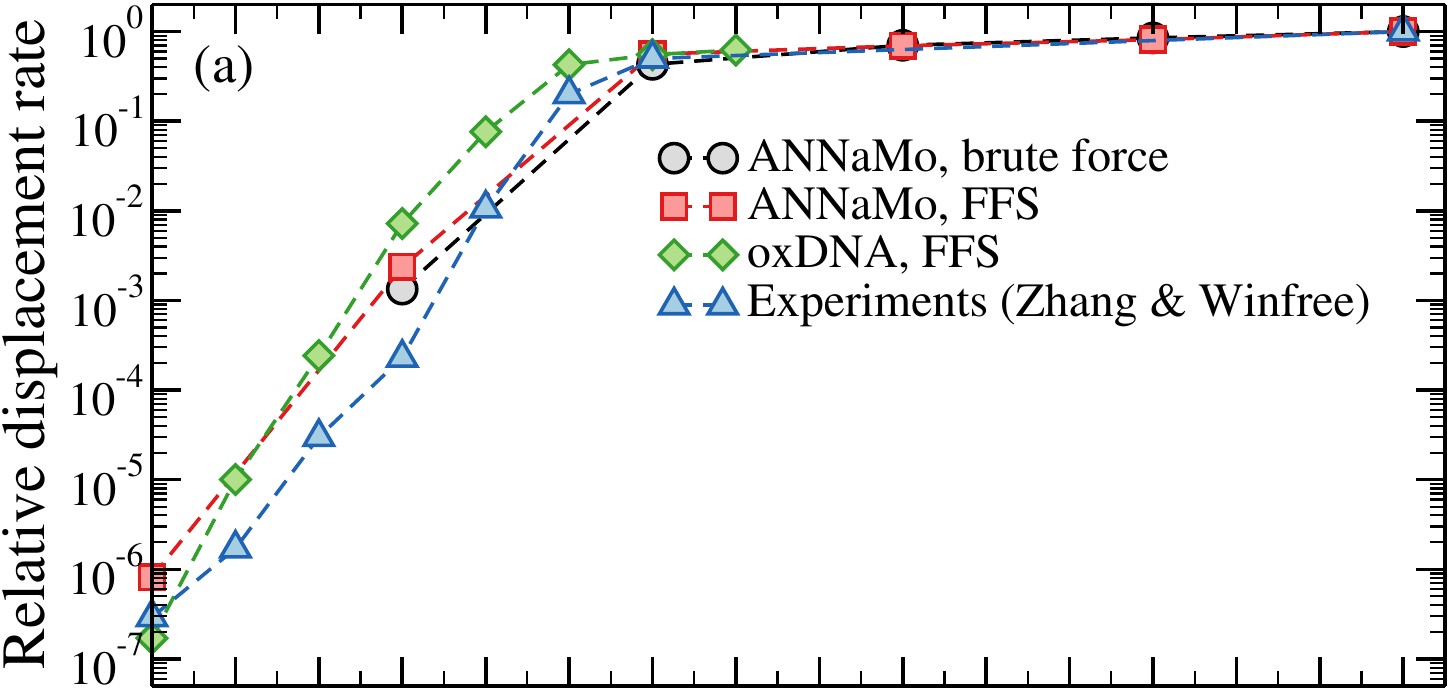}\\
      \includegraphics[width=.45\textwidth]{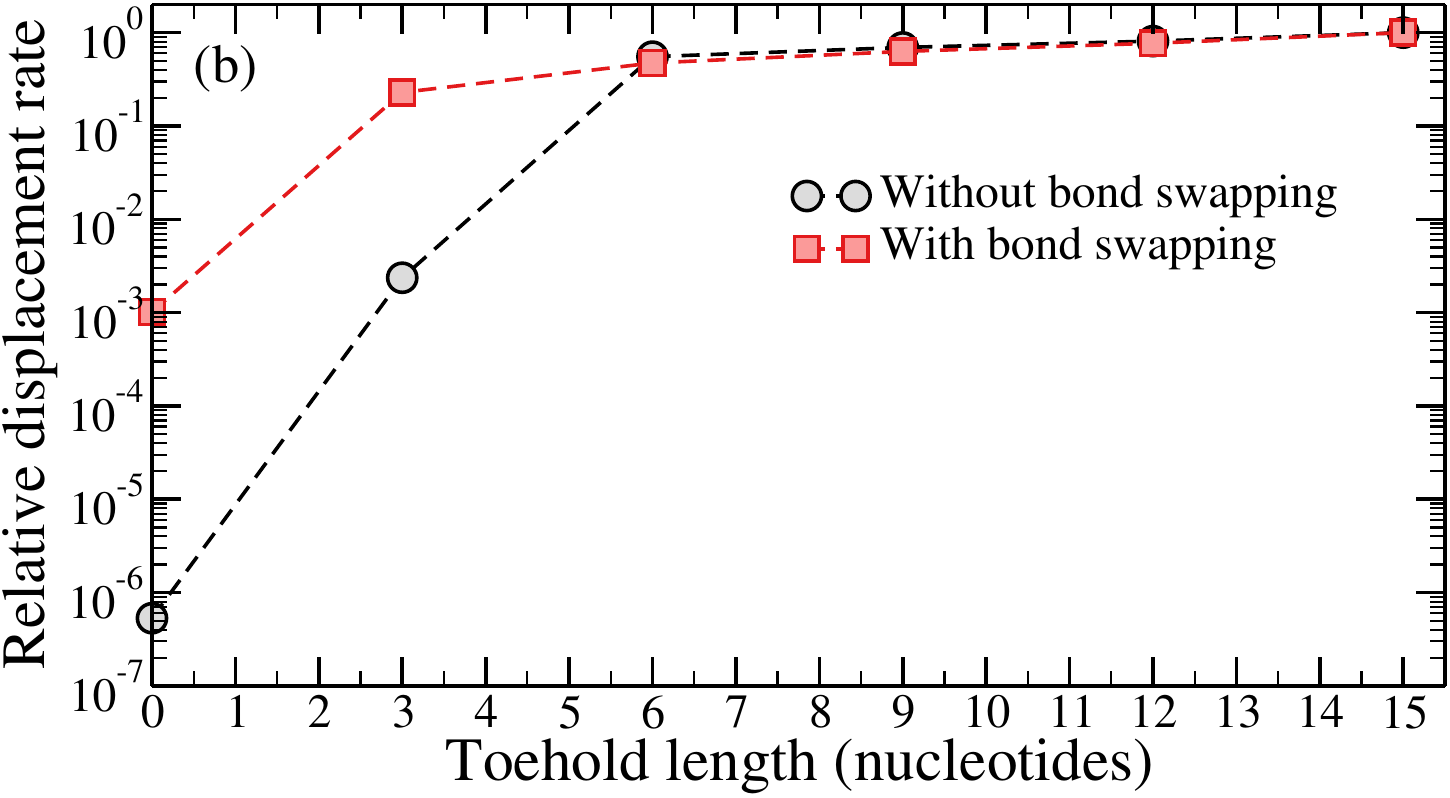}
\caption{(a) Displacement rates as a function of toehold length (in number of nucleotides) as computed in simulations (\textit{via} brute-force or FFS calculations) and experiments~\cite{zhang2009control}, relative to the longest-toehold case (7 nucleotides for the oxDNA data, 15 nucleotides in all other cases). (b) The same quantity evaluated with \model{}, with and without bond swapping. Errors are always smaller than symbol size.}\label{fig: toehold}
\end{figure}

Figure~\ref{fig: toehold} shows the relative displacement rates as a function of toehold length as obtained with \model{} and compared with experiments and oxDNA simulations (Fig.~\ref{fig: toehold}a), as well as a comparison between the relative displacement rates with and without bond swapping (Fig.~\ref{fig: toehold}b).

It is striking to note that the \model{} results are at least as good as those obtained with oxDNA, but at a fraction of the computational cost:  with the new model, the average (single-CPU-core) walltime required by unbiased MD simulations to observe displacement in systems with toeholds longer than one bead is smaller than one minute at a strand concentration of $\approx 1$~mM. However, for shorter toeholds even with \model{} TMSD processes become hard to probe due to the exponential dependence of the rate and, unless bond-swapping is enabled, FFS or equivalent techniques are needed. However, as shown in Fig.~\ref{fig: toehold}b, if bond-swapping is enabled, the time-dependence of the displacement rate is weaker, and displacements in one-bead-long toehold are only marginally slower than in the case of longer toeholds. Such a high sampling efficiency makes it possible to investigate, and possibly optimise the kinetics of, more complicated systems featuring many different strand displacement gates \cite{qian2011neural}.

\section{Discussion and conclusions}

This study presents a novel coarse-grained model aimed at simulating folding processes of DNA and RNA nanostructures, \model{}. By representing $n = 3$ nucleotides with a single patchy particle, we have achieved a balance between computational efficiency and the ability to capture the interactions that govern folding dynamics, as demonstrated by simulations of DNA hairpins, an RNA pseudoknot, and an RNA tile. The model is parametrized using well-established nearest-neighbor models, and can offer insights into the stability and thermodynamics of nucleic acid structures, while opening avenues for the exploration of larger systems and longer timescales. 

We showed that the thermodynamic performance of the model is comparable with those of the best nucleotide-level coarse-grained models (oxDNA and oxRNA), while being two orders of magnitude faster. From the kinetic point of view, \model{} reproduces the dependence of the rate of toehold-mediated strand displacement processes on toehold length as observed with experiments or nucleotide-level simulations.

While the model focuses on thermodynamics, some structural properties of nucleic acids, such as the different persistence length between single- and double-stranded molecules, are also retained. Finally, we showed that with \model{} it is possible to straightforwardly obtain melting curves of larger structures, such as an RNA tile, \textit{via} unbiased simulations. However, in this case we have observed a likely overestimation of the stability of some specific motifs, such as kissing loops.

Possible future applications of the model are the exploration of folding pathways, optimization of single-stranded motifs and origami designs~\cite{han2017single,qi2018programming}, vaccine design~\cite{leppek2022combinatorial,KIM2023}, and viral RNA folding and packaging~\cite{LosdorferBozic_2018,10.1063/5.0152604}. Although the possibility of simulating multi-stranded systems makes it possible to simulate the formation of complex nanostructures such as DNA origami with the aim of understanding and optimising their folding pathways, doing so may require adding additional (coaxial) stacking interactions between the domains~\cite{D3NR02537C,doi:10.1021/acs.nanolett.2c01372}.

\begin{acknowledgments}
We thank Francesco Sciortino for the fruitful discussions. P\v{S} acknowledges support by the National Science Foundation under Grant DMR-2239518.
\end{acknowledgments}

\appendix

\section{The functional forms of the interaction potential}
\label{app:model}
\begin{figure}
\centering
\includegraphics[width=0.45\textwidth]{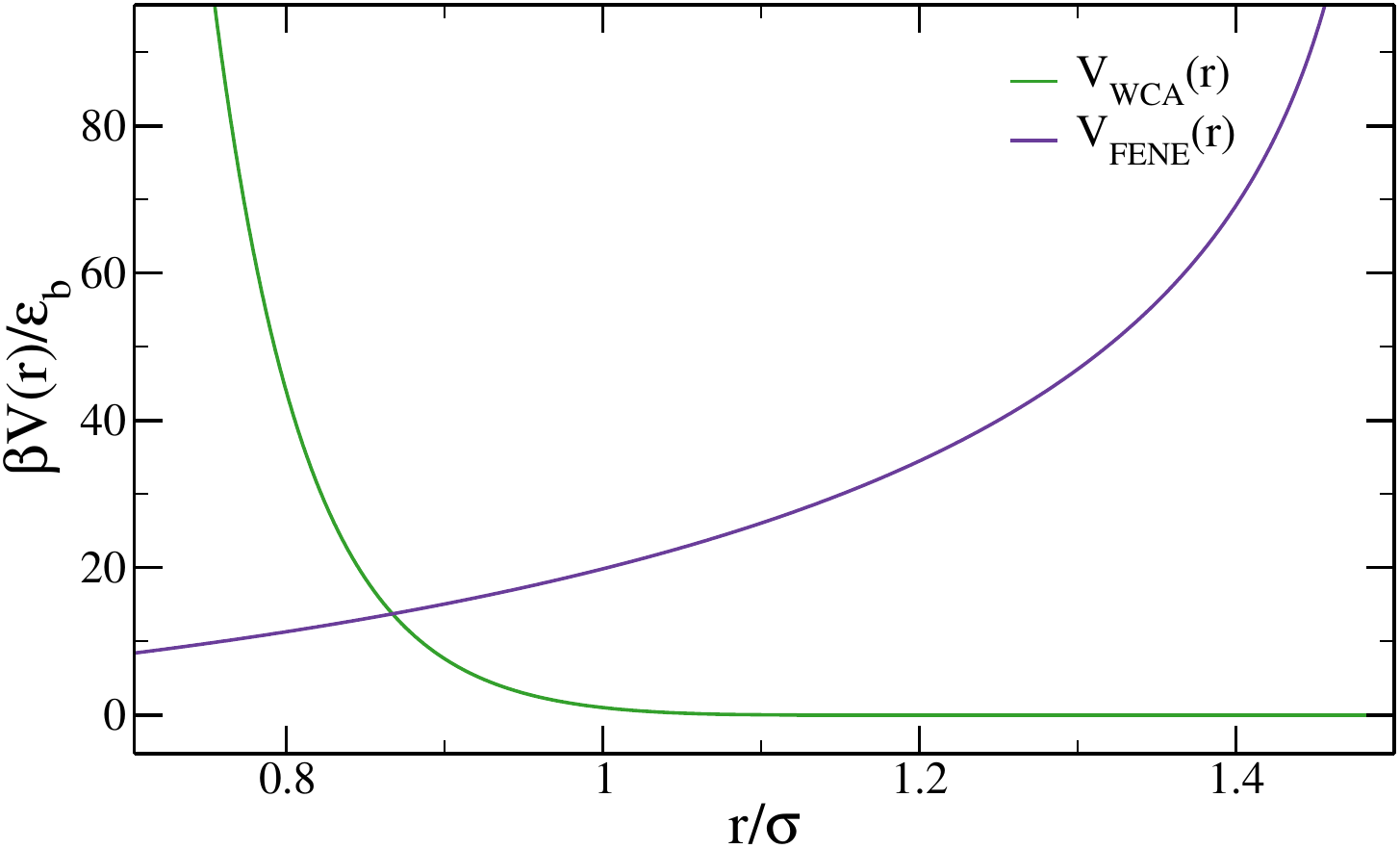}
\caption{The different potentials acting between bonded beads.}
\label{fig: BondedPotentials}
\end{figure}

\begin{figure}
    \centering
    \includegraphics[width=0.45\textwidth]{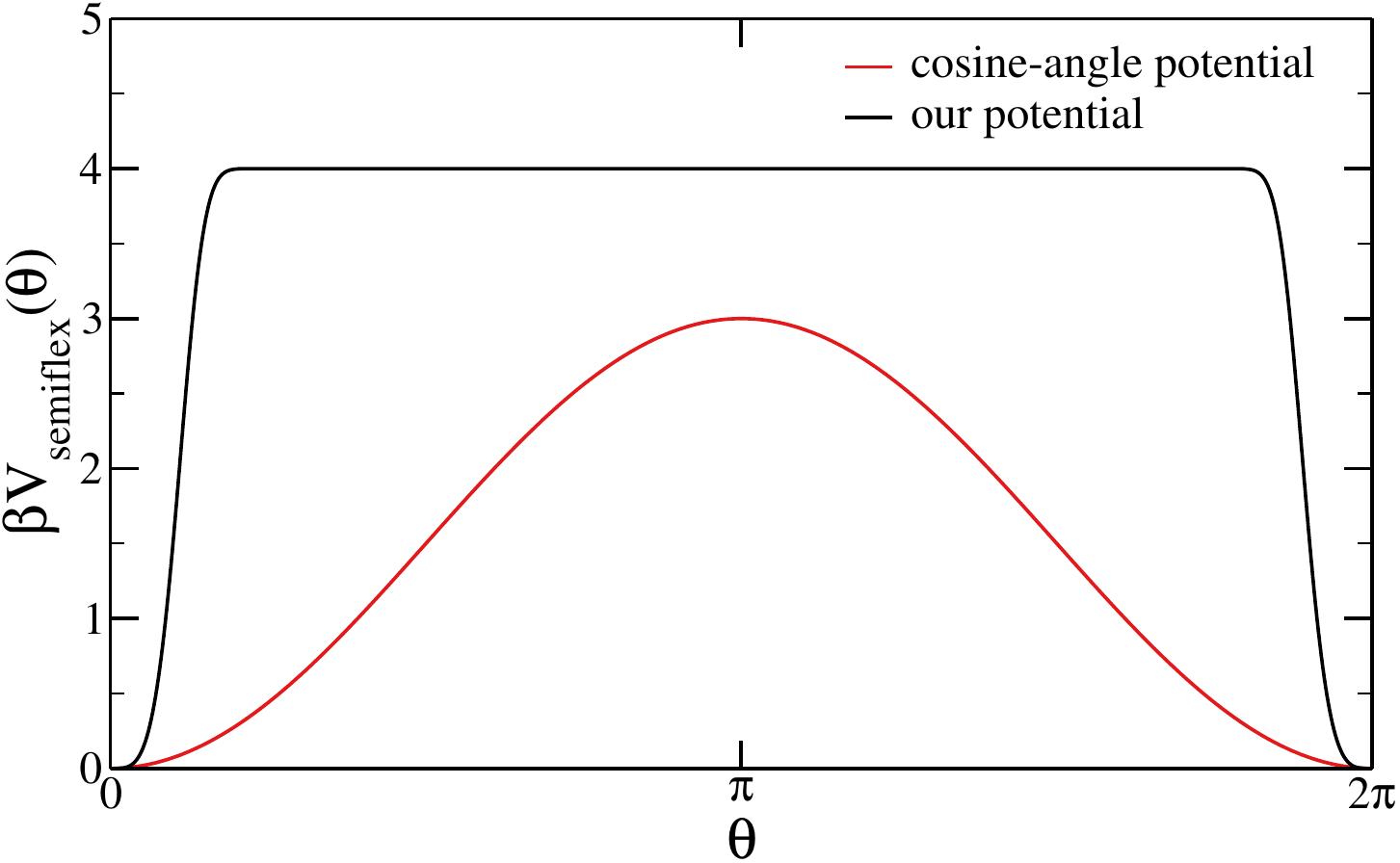}
    \caption{Plot depicting a potential designed to simulate the distinct behaviors of DNA/RNA in single- and double-stranded configurations (black curve). A widely used cosine-angle potential to introduce structural rigidity is included for comparison (red curve). The advantage of our potential is that when there are fewer constraints like in the single-stranded case, the polymer is more likely to rotate freely.}
    \label{fig: lr_pot}
\end{figure}

The potentials between \textbf{topologically bonded beads} (beads linked through the backbone) are:
\begin{itemize}
\item The Kremer-Grest force field \cite{PhysRevA.33.3628} $V_{KG}$, which is the sum of a WCA potential and a FENE potential (see Figure \ref{fig: BondedPotentials}). This spring-like potential guarantees excluded volume (WCA component) and attraction (FENE component) that mimic the covalent bonds along the strand.
    In particular, defining $r$ as the distance between bonded beads, the WCA potential is
    \begin{equation}
    \label{eq:WCA}
    V_{WCA}(r) = 
    \begin{cases}
      4\varepsilon \left[ \left( \frac{\sigma}{r} \right)^{12} -  \left( \frac{\sigma}{r} \right)^{6} \right] + \varepsilon, &\, r\leq 2^{1/6}\sigma \\
      0, & \, r> 2^{1/6}\sigma
    \end{cases}
    \end{equation}
    where $\varepsilon$ and $\sigma$ are the units of energy and length respectively (both set to 1 in our simulations),
    while the FENE potential is
    \begin{equation}
    V_{FENE}(r) = -\frac{1}{2} K d_0^2 \ln \left[1-\left( \frac{r}{d_0} \right)^2 \right],
    \end{equation}
    where $d_0=1.5\sigma$ and $K=30\varepsilon /\sigma^2$. 
\item A three-body potential that tends to align three consecutive beads (see Figure \ref{fig: lr_pot}):
    \begin{equation}
        V_{\rm semiflex}(\theta) = -k \left[ e^{-\left(\frac{1-\cos{\theta}}{\xi}\right)^2} -1 \right] .
    \end{equation}
    \noindent where $\theta$ is the angle defined by a triplet of bonded particles.
\item A term $V_{\rm stack}$ that models the stacking of base in DNA/RNA, acting on the $\Vec{a}_1$ directions of consecutive beads:
    \begin{equation}
        V_{\rm stack} (\Vec{a}_1^{(i)}, \Vec{a}_1^{(i+1)}) = \eta (1 - \Vec{a}_1^{(i)} \cdot \Vec{a}_1^{(i+1)}).
    \end{equation}
\end{itemize}

\textbf{Non-bonded beads} interact through the following potentials:

\begin{itemize}
\item A WCA potential, Eq.~\eqref{eq:WCA}, to model the excluded-volume interaction;
\item An attractive potential that models the hybridization of nucleotides, $V_{\rm sticky} \cdot V_{\rm direct}$. Each patch has a position that is given by the position of the bead it is attached to plus $\delta_{pb} \vec{a}_1$, where $\delta_{pb}=0.65\sigma$, and $V_{\rm sticky}(r_{pp})$ acts between any two patches, Its  functional form was proposed by Stillinger and Weber \cite{PhysRevB.31.1954} (see Figure \ref{fig: potentials}) and reads
    \begin{equation}
        V_{\rm sticky}^{ij}(r_{pp}) = A\varepsilon_{ij} \left[ B \left( \frac{\sigma_s}{r_{pp}} \right)^4 -1 \right] e^{\sigma_s/(r_{pp}-r_c)}    
    \end{equation}
    where $i$ and $j$ are the bead types and $r_{pp}$ is the distance between the patches and
    \begin{equation}
    \begin{split}
    B &= \frac{1}{1+4\left( 1-r_s\right)^2},  \\
    A &= -\frac{1}{B-1}\frac{1}{e^{1/1-r_s}},  \\
    r_c &= \sigma_s r_s.
    \end{split}
    \end{equation}
    The coefficient $\varepsilon_{ij}$ modulates the strength of the sticky attraction and depends on what nucleotides are inside the considered beads (see Section~\ref{Parametrization}). We set $\varepsilon_{ij} = 0$ between beads $i$ and $j$ that are first neighbors since loops with lengths shorter than 3 are sterically prohibited \cite{doi:10.1146/annurev.biophys.32.110601.141800}.
    This potential depends essentially on two parameters: $\sigma_s$, which defines the minimum of the potential (\textit{i.e.} the radius of the patch) and is set to $0.21875\sigma$, and $r_s$, which defines the steepness of the potential between the minimum and $r_c$ (set to $0.35$), after which $V_{\rm sticky}=0$.
    Thus, two particles are bonded if the relative distance between their patches is less than $r_c$. The directionality of DNA/RNA is enforced by multiplying $V_{\rm sticky}$ by a term acting on the $\Vec{a}_3$ directions of beads:
    \begin{equation}
        V_{\rm direct} (\Vec{a}_3^{(i)}, \Vec{a}_3^{(j)}) = \frac{\Vec{a}_3^{(i)} \cdot \Vec{a}_3^{(j)}}{2}.
    \end{equation}
    This term makes sure that only antiparallel strands can bind to each other.
\item $V_{\rm 3b}$, To ensure the single-bond-per-bead condition, we implement a repulsive three-body interaction $V_{\rm 3b}$ which penalizes the formation of triplets of bonded beads \cite{Sciortino2017}. In particular, $V_{\rm 3b}$ is designed to almost exactly compensate the gain associated with the formation of a second bond, originating an almost flat energy hypersurface that favors bond swapping even when the bonding energy is much larger than the thermal energy. This repulsive potential is defined as
    \begin{equation}
    V_{\rm 3b} = \lambda  \sum_{i j k} \min(\varepsilon_{ij},\varepsilon_{ik}) V_3(r_{ij}) V_3(r_{ik})
    \end{equation}
where the sum runs over all triplets of bonded particles (bead $i$ bonded both with $k$ and $j$), $r_{ij}$ is the distance between bead $i$ and $j$, and the minimum between $\varepsilon_{ij}$ and $\varepsilon_{ik}$ is chosen to favor the removal of the more loosely bonded bead. The value of the parameter $\lambda$ allows to interpolate between the limits of swapping ($\lambda = 1$) and non-swapping ($\lambda \gg  1$) bonds. The pair potential $V_3(r)$ is defined in terms of the normalized $V_{\rm sticky}(r)$ as
    \begin{equation}
    V_3(r)=
    \begin{cases}
      1, &\qquad r\leq \sigma_s \\
      -\frac{V_{\rm sticky}^{ij}(r)}{\varepsilon_{ij}}, &\qquad \sigma_s\leq r \leq r_c
    \end{cases}
    \label{eq: V3}
    \end{equation}
where $\sigma_{s}$ is the distance at which $V_{sticky}^{ij}(r)$ has a minimum.
    \begin{figure}
    \centering
    \includegraphics[scale =0.3]{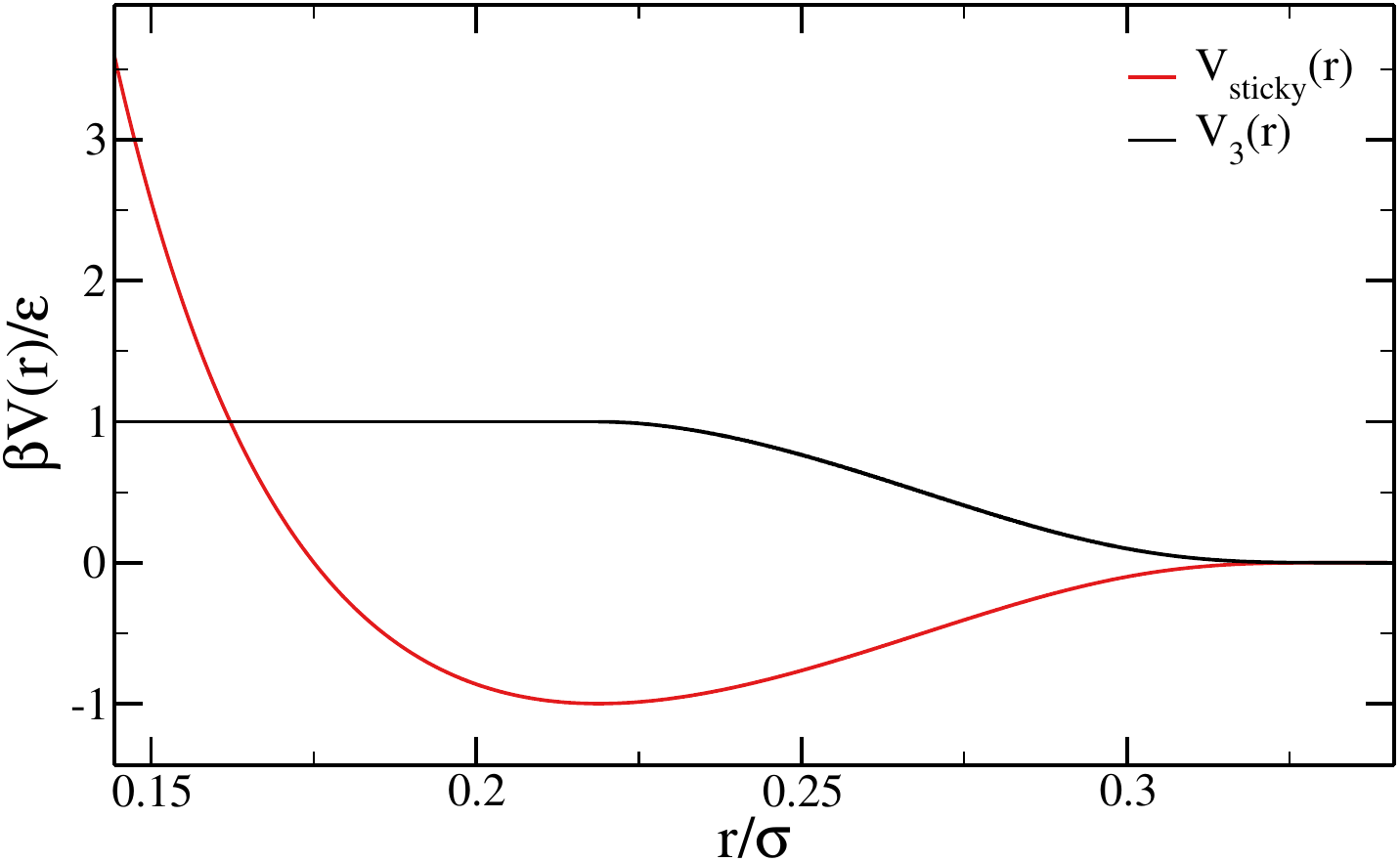}
    \caption{The different potentials acting between non-bonded beads.}
    \label{fig: potentials}
    \end{figure}
\end{itemize}

\section{Mapping}
\label{Mapping}

Here we establish a connection between \model{} and the NN models by using the parameters of the latter to set the strength of the sticky interaction of the former. In order to do so we adopt the procedure of Ref.\cite{reinhardt2016dna}: the equilibrium constant associated with the chemical equilibrium of a dimer $AB$ and two beads $A$ and $B$ is given by:
\begin{equation}\label{eq: equilibrium}
    K = \frac{[AB]/[\stkout{\circ}]}{([A]/[\stkout{\circ}])([B]/[\stkout{\circ}])}=\frac{\rho_{AB}\rho^{\stkout{\circ}}}{\rho_A\rho_B}=\exp{(-\beta\Delta G^{\stkout{\circ}})},
\end{equation}
where $[\stkout{\circ}]=1 \,\textrm{mol dm}^{-3}$ is the standard state concentration, $\rho^{\stkout{\circ}}=[\stkout{\circ}]\mathcal{N}_A=6.022\times 10^{26}\,\textrm{m}^{-3}$ is the standard number density and $\Delta G^{\stkout{\circ}}$ is the standard Gibbs energy for the transformation $A + B \rightleftharpoons AB$ where 50\% of the beads have hybridized.
The right-hand side of eq.\ref{eq: equilibrium} can be written as 
\begin{equation}\label{eq: q_ab}
    q_{AB}\rho^{\stkout{\circ}}=\exp{(-\beta \Delta G^{\stkout{\circ}})}.
\end{equation}
where $q_{AB}$ is the internal partition function of $AB$.
Knowing that:
\begin{equation}\label{eq: integral}
    q_{AB} = 4\pi \int r^2dr \int d\{\Vec{a}\} \int d \{\Vec{b}\} e^{-\beta V_{\rm tot}}
\end{equation}
where $V_{\rm tot}$ is the total interaction potential between the two beads, we evaluate the right-hand side of Eq.~\ref{eq: integral} through a Monte Carlo integration, finding that, for $\varepsilon \gtrsim 3$, $q_{AB} \simeq V_b e^{\alpha \beta \varepsilon}$, where $V_b = 0.0019 \, \sigma^3$ and $\alpha = 0.89$ are fitting parameters. Substituting this relation in eq.\ref{eq: q_ab} we find

\begin{equation}
    \Tilde{V}_be^{\alpha \beta \varepsilon_{ij}}\sigma^3 \rho^{\stkout{\circ}} = e^{-\beta \Delta G^{\stkout{\circ}}},
\end{equation}

\noindent so that

\begin{align}
    \ln{(\rho^{\stkout{\circ}}\sigma^3\Tilde{V}_b)} + \alpha\beta\varepsilon_{ij} &= -\beta\Delta G^{\stkout{\circ}} \notag \\
    \beta \varepsilon_{ij} &= -\frac{1}{\alpha} \left[ \beta \Delta G^{\stkout{\circ}} + \ln{(\rho^{\stkout{\circ}}\sigma^3\Tilde{V}_b)} \right]
\end{align}
Since $\Delta G = \Delta H - T\Delta S$, we can rewrite the previous equation as:
\begin{equation}
    \beta \varepsilon_{ij} = -\frac{1}{\alpha} \left[ \frac{1}{k_BT} (\Delta H - T\Delta S) + \ln{(\rho^{\stkout{\circ}}\sigma^3\Tilde{V}_b)} \right].
\end{equation}
Finally, using eq.\ref{eq: dG def}, we obtain:
\begin{equation}
    \beta \varepsilon_{ij} = -\frac{1}{\alpha}\left[ \left(\frac{\Delta H}{k_BT_{37}} \frac{T_{37}}{T} -\frac{\Delta S}{k_B} \right) +  \ln{(\rho^{\stkout{\circ}}\sigma^3\Tilde{V}_b)} \right].
\end{equation}

\section{Hairpin sequences}
\label{sequences}

The sequences used for studying the melting curves of the hairpins \ref{sub: DNA hairpins} are listed in Table \ref{tab:hair6_seq} and Table \ref{tab:hair9_seq}.
\input{tables/seq_loop6}
\input{tables/seq_loop9}

\nocite{*}
\bibliography{biblio}

\end{document}

%% file: tables/6stem6loop.tex
    \centering
    \begin{tabular}{cccc}
    \hline
        6stem6loop & oxDNA & ANNaMo & $\Delta$ \\ \hline
        DNAh1 & 66 & 67 & 1 \\ \hline
        DNAh2 & 72 & 72 & 0 \\ \hline
        DNAh3 & 64 & 66 & 2 \\ \hline
        DNAh4 & 61 & 59 & -2 \\ \hline
        DNAh5 & 64 & 67 & 3 \\ \hline
        DNAh6 & 64 & 62 & -2 \\ \hline
        DNAh7 & 68 & 64 & -4 \\ \hline
        DNAh8 & 71 & 71 & 0 \\ \hline
        DNAh9 & 63 & 66 & 3 \\ \hline
        DNAh10 & 60 & 59 & -1 \\ \hline
        DNAh11 & 65 & 64 & -1 \\ \hline
        DNAh12 & 65 & 68 & 3 \\ \hline
        ~ & ~ & ~ & ~ \\ \hline
        ~ & ~ & ~ & $\langle|\Delta|\rangle$ \\ \hline
        ~ & ~ & ~ & 1.8 \\ \hline
    \end{tabular}

%% file: tables/6stem9loop.tex
    \centering
    \begin{tabular}{cccc}
    \hline
        6stem9loop & oxDNA & ANNaMo & $\Delta$ \\ \hline
        DNAh1 & 62 & 65 & 3 \\ \hline
        DNAh2 & 66 & 68 & 2 \\ \hline
        DNAh3 & 60 & 62 & 2 \\ \hline
        DNAh4 & 57 & 56 & -1 \\ \hline
        DNAh5 & 60 & 64 & 4 \\ \hline
        DNAh6 & 59 & 60 & 1 \\ \hline
        DNAh7 & 61 & 62 & 1 \\ \hline
        DNAh8 & 65 & 67 & 2 \\ \hline
        DNAh9 & 59 & 63 & 4 \\ \hline
        DNAh10 & 56 & 56 & 0 \\ \hline
        DNAh11 & 59 & 62 & 3 \\ \hline
        DNAh12 & 60 & 65 & 5 \\ \hline
        ~ & ~ & ~ & ~ \\ \hline
        ~ & ~ & ~ & $\langle|\Delta|\rangle$ \\ \hline
        ~ & ~ & ~ & 2.3 \\ \hline
    \end{tabular}

%% file: tables/seq_loop6.tex
\begin{table}[!ht]
    \centering
    \begin{tabular}{c c}
    \hline
        6stem6loop  & \\ \hline
        DNAh1:5'-GCGTTGCTTCTCCAACGC-3' & \\ \hline
        DNAh2:5'-TTGGCGCTTCTCCGCCAA-3' & \\ \hline
        DNAh3:5'-AGGCTCCTTCTCGAGCCT-3' & \\ \hline
        DNAh4:5'-CTCAGGCTTCTCCCTGAG-3' & \\ \hline
        DNAh5:5'-GGACGTCTTCTCACGTCC-3' & \\ \hline
        DNAh6:5'-CGTGGACTTCTCTCCACG-3' & \\ \hline
        DNAh7:5'-CGCAACCTCTTCGTTGCG-3' & \\ \hline
        DNAh8:5'-AACCGCCTCTTCGCGGTT-3' & \\ \hline
        DNAh9:5'-TCCGAGCTCTTCCTCGGA-3' & \\ \hline
        DNAh10: 5'-GAGTCCCTCTTCGGACTC-3' & \\ \hline
        DNAh11: 5'-CCTGCACTCTTCTGCAGG-3' & \\ \hline
        DNAh12: 5'-GCACCTCTCTTCAGGTGC-3' & \\ \hline

    \end{tabular}
    \caption{DNA sequences used for studying the melting curves of hairpins with a stem length of six and a loop length of six.}\label{tab:hair6_seq}
\end{table}

%% file: tables/seq_loop9.tex
\begin{table}[!ht]
    \centering
    \begin{tabular}{c c}
    \hline
        6stem9loop  & ~ \\ \hline
        DNAh1:5'-GCGTTGCTATGCTTCCAACGC-3' & ~ \\ \hline
        DNAh2:5'-TTGGCGCTATGCTTCCGCCAA-3' & ~ \\ \hline
        DNAh3:5'-AGGCTCCTATGCTTCGAGCCT-3' & ~ \\ \hline
        DNAh4:5'-CTCAGGCTATGCTTCCCTGAG-3' & ~ \\ \hline
        DNAh5:5'-GGACGTCTATGCTTCACGTCC-3' & ~ \\ \hline
        DNAh6:5'-CGTGGACTATGCTTCTCCACG-3' & ~ \\ \hline
        DNAh7:5'-CGCAACCTACGTTTCGTTGCG-3' & ~ \\ \hline
        DNAh8:5'-AACCGCCTACGTTTCGCGGTT-3' & ~ \\ \hline
        DNAh9:5'-TCCGAGCTACGTTTCCTCGGA-3' & ~ \\ \hline
        DNAh10: 5'-GAGTCCCTACGTTTCGGACTC-3' & ~ \\ \hline
        DNAh11: 5'-CCTGCACTACGTTTCTGCAGG-3' & ~ \\ \hline
        DNAh12: 5'-GCACCTCTACGTTTCAGGTGC-3' & ~ \\ \hline

    \end{tabular}
    \caption{DNA sequences used for studying the melting curves of hairpins with a stem length of six and a loop length of nine.}\label{tab:hair9_seq}
\end{table}